\documentclass[traditabstract]{aa}
\usepackage{natbib}
\usepackage{fancyhdr}
\usepackage{graphicx}
\usepackage{float}
\usepackage{subfigure}
\usepackage{amsmath}
\usepackage{amssymb}
\usepackage{amsfonts}
\usepackage{txfonts}
\usepackage{aas_macros}
\usepackage{dcolumn}
\usepackage{dsfont}

\usepackage{multirow}

\newcommand{\pa}[2] {\frac{\partial #1}{\partial #2}}
\newcommand{\totdiff}[2]{\frac{\text{d} #1}{\text{d} #2} }
\newcommand{\cms}{\text{ cm} \cdot \text{s}^{-1}}

\renewcommand{\vec}[1]{{\boldsymbol{#1}}}

\usepackage{natbib,twoopt}
 \usepackage[breaklinks=true]{hyperref} %% to avoid \citeads line fills
 \bibpunct{(}{)}{;}{a}{}{,}    %% natbib format like A&A and ApJ
 \newcommandtwoopt{\citeads}[3][][]{\href{http://adsabs.harvard.edu/abs/#3}%
                                        {\citealp[#1][#2]{#3}}}
 \newcommandtwoopt{\citepads}[3][][]{\href{http://adsabs.harvard.edu/abs/#3}%
                                        {\citep[#1][#2]{#3}}}
 \newcommandtwoopt{\citetads}[3][][]{\href{http://adsabs.harvard.edu/abs/#3}%
                                        {\citet[#1][#2]{#3}}}
 \newcommandtwoopt{\citeyearads}[3][][]%
   {\href{http://adsabs.harvard.edu/abs/#3}{\citeyear[#1][#2]{#3}}}

\pdfmapfile{+txfonts.map}

\begin{document}

\title{Particle scattering in turbulent plasmas with amplified wave modes}

\author{Sebastian Lange\inst{\ref{inst1}} \and Felix Spanier\inst{\ref{inst2}} \and Markus Battarbee\inst{\ref{inst3}}\and Rami Vainio\inst{\ref{inst4}}  \and Timo Laitinen\inst{\ref{inst5}}}

\institute{Lehrstuhl f\"ur Astronomie, Universit\"at W\"urzburg, Emil-Fischer-Stra\ss e 31, D-97074 W\"urzburg \label{inst1} \and Centre for Space Research, North-West University, 2520 Potchefstroom, South Africa \label{inst2} \and Department of Physics and Astronomy, University of Turku, 20014 Finland \label{inst3} \and Department of Physics, P.O.Box 64, 00014 University of Helsinki, Finland\label{inst4}  \and Jeremiah Horrocks Institute, University of Central Lancashire, PR1 2HE, Preston, UK \label{inst5} }
\date{}

\abstract{High-energy particles stream during coronal mass ejections or flares
through the plasma of the solar wind. This causes instabilities, which lead to
wave growth at specific resonant wave numbers, especially within shock regions.
These amplified wave modes influence the turbulent scattering process
significantly. In this paper, results of particle transport and scattering in
turbulent plasmas with excited wave modes are presented. The method used is a
hybrid simulation code, which treats the heliospheric turbulence by an
incompressible magnetohydrodynamic approach separately from a kinetic particle
description. Furthermore, a semi-analytical model using quasilinear theory (QLT) is
compared to the numerical results. This paper aims at a more fundamental
understanding and interpretation of the pitch-angle scattering coefficients. \\
Our calculations show a good agreement of particle simulations and the QLT for
broad-band turbulent spectra; for higher turbulence levels and particle beam
driven plasmas, the QLT approximation gets worse. Especially the resonance gap
at $\mu=0$ poses a well-known problem for QLT for steep turbulence spectra, 
whereas test-particle computations show no problems for the particles to scatter across this region. 
The reason is that the sharp resonant wave--particle interactions in QLT are an oversimplification 
of the broader resonances in test-particle calculations, which result from nonlinear effects not included in the QLT. 
We emphasise the importance of these results for both numerical simulations and analytical particle transport
approaches, especially the validity of the QLT.  
} 


\titlerunning{Particle scattering at peaked modes}
\authorrunning{S. Lange et al.}

\maketitle

\section{Introduction}\label{sec:intro}

Coronal mass ejections (CME) are believed to be an important source of solar energetic particles (SEP). 
The acceleration mechanism taking place is the diffusive shock acceleration process \citepads{reames99}. 
One cornerstone in the theory behind this mechanism is understanding particle scattering of SEP in the background plasma. 
The general approach used for this purpose is the quasilinear theory (QLT) \citepads{jokipii66}.

The CME may be represented as a system of a thermal plasma with a turbulent energy spectrum and a nonthermal proton and electron component. 
These constituents are interacting with each other in a nonlinear way. 
In order to understand this system, it is necessary to model wave excitation by particle beams, turbulent energy transfer, and the transport of high-energy particles. 
Models have been developed by, e.g. \citetads{2007ApJ...658..622V} and \citetads{2008ApJ...686L.123N}. 
While these models are capable of describing the interaction of particles and waves, the details of the specific processes are implemented in a simple way,
basically relying on QLT or its simple extensions as the correct description of particle scattering and ignoring oblique or perpendicular
wave modes completely. 
\citetads{lange2012} did numerical simulations of the turbulent cascading process of energy injected via proton beams. 
The basic concept of this paper is to take first steps towards a consistent description of particle transport within these plasma configurations and
investigate how well the QLT describes scattering in such turbulence.

The general problem in the derivation of transport coefficients is a gap between theoretical and numerical approaches. Numerical simulations have used artificial turbulence spectra \citepads{2002ApJ...578L.117Q}, while analytical calculations use any possible spectrum. However, the applicability of these calculations is unclear. 
This work combines the previous turbulence simulations by \citetads{lange2012} with a particle tracking algorithm derived from \citetads{2011ASTRA...7...21S}. 
The results are compared with QLT calculations using an identical input spectrum.

This approach allows a comparison of QLT and numerical simulations. The latter are assumed to be valid for a wide range of turbulent spectra and particle energies. Therefore, a determination of the validity of QLT is possible. 
Novel visualization and comparison methods are used to probe nonlinear and time-dependent scattering processes. Thus, a more fundamental understanding of the scattering process is achieved.

\section{Theory}\label{sec:theory}

\subsection{Numerical Model}\label{sec:numerics}

The region of the heliosphere we are interested in is within the weak turbulence regime, with magnetic field fluctuations defined as
\begin{align}
 \delta \vec{B} \equiv \vec{B} - \vec{B}_0,
\end{align}
where $\langle \delta \vec{B} \rangle = 0$  is the mean value of the fluctuations, which leads to  $\langle \vec{B} \rangle = \vec B_0$. 
In that sense, weak turbulence is connected to small fluctuation amplitudes compared to the magnetic background field $B_0$. In terms of wave turbulence, this means that the linear Alfv\'en wave timescale is shorter than the intrinsically nonlinear time.
It is observed that the solar wind magnetic fluctuations decrease as $\delta B^2\propto r^{-3}$, while the background field decreases as $B_0^2\propto r^{-4}$
\citepads{1982SoPh...78..373B,brunocarbone-livrev}. Consequently the $\delta B/B_0$ ratio within the heliosphere is increasing with distance from the Sun
\citepads{hollweg10}.

The magnetic background field $\vec{B}_0$ is defined towards the z-direction within our simulations and hence is also denoted as $B_0 \vec{e}_z$. The parallel direction is, therefore, defined as the z-direction and the x- and y-direction are the perpendicular directions. For symmetry reasons, there will be no further distinction between the two perpendicular directions, and all plots show values averaged over the azimuthal angle in cylindrical coordinates of the x--y plane.

High Reynolds numbers in combination with massive energy injection, as seen in, e.g. the solar wind, are strong indicators of a highly turbulent state. 
This means that most parts of the heliosphere are dominated by turbulent evolution and consequently energy cascading to smaller spatial scales.
In situ measurements of the energy spectrum \citepads{tu-marsch} are in agreement with this fact.

To simulate conditions within the turbulent heliospheric plasma with particle transport and scattering, the research group at the University of W\"urzburg has developed a hybrid simulation code, \textsc{Gismo}.
It is divided into two parts, the magnetohydrodynamic (MHD) algorithm \textsc{Gismo--MHD} and the kinetic particle transport simulation tool \textsc{Gismo--Particles}.

\textsc{Gismo--MHD} is an incompressible pseudospectral MHD code that is fully parallelised and capable of efficiently using massive computing clusters. For a more detailed description we refer to \citetads{lange2012}. A brief overview is presented below.
The basis of the simulation software is to solve the following set of incompressible MHD equations:
\begin{align}
 \pa{\vec{u}}{t} &= \vec{b} \cdot \nabla \vec{b} -\vec{u} \cdot \nabla \vec{u} -\nabla P + \nu_v \nabla^{2h} \vec{u}  \nonumber \\
 \pa{\vec{b}}{t} &= \vec{b} \cdot \nabla \vec{u} -\vec{u} \cdot \nabla \vec{b} + \eta \nabla^{2h} \vec{b} \nonumber \\
 \nabla \cdot \vec{u} &= 0 \nonumber \\
 \nabla \cdot \vec{b} &= 0, \label{eq:mhdset}
\end{align}
where $\vec{b}={\vec{B}}/{\sqrt{4\pi \varrho}}$ is the normalised magnetic field, $\vec{u}$ is the fluid velocity, and $\varrho$ is the constant mass density. The dissipative terms due to viscous and Ohmic dissipation are denoted by $\nu_v$ and $\eta$. In the following we introduce the parameter $\nu$ as a global diffusivity with $\eta=\nu_v\equiv\nu$. The approach is valid for magnetic Prandtl number of order unity, which is applicable within the regime of Alfv\'en wave turbulence where an equipartition between magnetic and kinetic energy can be assumed \citepads{PhysRevE_66_046410,2008AA...490..325B}.
The artificial increase of the dissipation by the hyperdiffusivity parameter $h$ is often used in spectral methods.
The pressure term $\nabla P$ fulfils the closure condition for incompressibility \citepads{marongold}:
\begin{align}
 \nabla^2 P = \nabla \vec{b} : \nabla \vec{b} -\nabla \vec{u} : \nabla \vec{u}. \label{eq:pressureclosure}
\end{align}

In the incompressible regime of a magnetised plasma, the MHD turbulence consists of only two types of waves, which transport energy along the parallel direction: the so--called pseudo and shear Alfv\'en waves. The former are the incompressible limit of slow magnetosonic waves and
play a minor role within anisotropic turbulence \citepads{marongold}. The pseudo Alfv\'en waves polarisation vector is in the plane
spanned by the wave vector $\vec{k}$ and $\vec{B}_0$.
The shear waves are transversal modes with  a polarisation vector perpendicular to the $\vec{k}$ - $\vec{B}_0$ plane. They are circularly polarised for parallel propagating waves.
Both species exhibit the dispersion relation $\omega^2=(v_A k_\parallel)^2$.
We note that the shear mode seems to be dominant because pseudo waves are heavily damped by the \emph{Barnes} damping process within weakly turbulent regimes \citepads{gsweak}. 
The damping weakens in strong turbulence, but according to \citetads{gsstrong}, the wave generation of pseudo Alfv\'enic wave modes is only possible via three-wave interactions by two shear wave modes. 
Barnes damping is important for high-$\beta$ plasmas. 
Since this is not fulfilled for the solar corona, the role of pseudo Alfv\'en waves should not be ignored. 
However, the solar wind and especially its fast component is strongly dominated by Alfv\'en waves \citepads{brunocarbone-livrev,2012ApJ...753L..19H} and therefore incompressible MHD is well applicable.

Because of this model, a suitable description of the system is achieved by the use of Alfv\'enic waves moving either forwards or backwards.
Therefore the Els\"asser variables \citepads{elsasser} are introduced:
\begin{align}
 \vec w^- &= \vec v + \vec b - v_A \vec e_\parallel \nonumber \\
 \vec w^+ &=\vec v - \vec b + v_A \vec e_\parallel.
\end{align}
Applying this definition to Eqs. (\ref{eq:mhdset}) and transformation to the Fourier space yields
\begin{align}
  \left(\partial_t - v_A k_z\right) \tilde w_\alpha^- &= \frac{i}{2} \frac{k_\alpha k_\beta k_\gamma}{k^2} \left( \widetilde{w_\beta^+ w_\gamma^-} +  \widetilde{w_\beta^- w_\gamma^+}\right) \nonumber \\
		                               &-ik_\beta \widetilde{w_\alpha^- w_\beta^+} - \frac{\nu}{2} k^{2h} \tilde w_\alpha^- \nonumber \\
  \left( \partial_t + v_A k_z \right) \tilde w_\alpha^+ &= \frac{i}{2} \frac{k_\alpha k_\beta k_\gamma}{k^2} \left(\widetilde{w_\beta^+ w_\gamma^-} +  \widetilde{w_\beta^- w_\gamma^+}\right) \nonumber \\
						&-ik_\beta \widetilde{w_\alpha^+ w_\beta^-} - \frac{\nu}{2} k^{2h} \tilde w_\alpha^+, \label{eq:fft-wpm-mhdset}
\end{align}
where the tilde--notation represents quantities in the Fourier space. This is the final set of equations which is iterated by \textsc{Gismo--MHD}.

Obviously, the nonlinearities of Eqs. (\ref{eq:fft-wpm-mhdset}) that describe the turbulent behaviour of the MHD plasma
cannot be solved in Fourier space in a straightforward fashion. Hence the main numerical load is the transfomation between real and wavenumber space for each
iterative step. For this purpose we used the P3DFFT algorithm, which is an MPI-parallelised fast Fourier transfomation based on FFTW3 \citepads{p3dfft}.

A basic problem of spectral methods that use discrete Fourier transformation is the aliasing effect. Because of discrete sampling in the wavenumber space,
high $k$-values exhibit errors that depend on the structure of the real space fields. Therefore we used zero padding, which is also referred to as Orszag's $2/3$
rule, i.e. $2/3$ of the wavenumbers below the Nyquist frequency have to be truncated to achieve maximum anti-aliasing, hence reducing the Fourier space resolution to $1/3$ of the original wavenumber range \citepads{orszag}. 
This process is repeated for each step, immediately before calculating the nonlinearities
and, accordingly, calculating the right-hand side (RHS) of the MHD equations. Consequently, the change in the antialiasing-range of one MHD step is physically correct, but not the long-term evolution.

The code \textsc{Gismo} is capable of using different foward--in--time schemes, namely, Euler and Runge-Kutta (RK) second as well as fourth order. All the simulations in this paper have been peformed using RK--4.

The Alfv\'en wave generation mechanism by SEPs is not investigated in detail here. The driving mechanism assumed for our simulations is the streaming
instability. The estimation of the wave growth rate is described in \citetads{rami-ongenofalfvs}. The streaming instability is caused by energetic proton
scattering off the coronal Alfv\'en waves. During the the scattering process the particle changes its pitch-angle cosine by $\Delta \mu$, while its
wave-frame energy remains constant. Thus the particles' energy in the plasma frame is changed by $v_A p \Delta \mu$, where $p$ is the wave-frame particle
momentum. Accordingly, the Alfv\'en wave energy in the plasma frame changes due to energy conservation. 
Another important instability in the solar corona is the electrostatic instability, which is caused by an electron current and streaming ions. 
Ion acoustic waves would be generated by this process. 
However, for the growth rate of these modes a sufficiently high ratio $T_e/T_i \gg1$ of the electron and ion temperatures is crucial.
Observations and simulations in the vicinity of three solar radii indicate temperature ratios of the order of 
unity \citepads{2007ApJ...663.1363L,2012ApJ...745....6J}. In this regime the ion acoustic waves are also efficiently suppressed by Landau damping. 
For further reading about the streaming instability we refer to \citetads{1993tspm.book.....G}.

The numerical tool for the kinetic simulation of charged test particles is \textsc{Gismo--Particles}. Like \textsc{Gismo--MHD} it is fully parallelised and capable of using massive computing clusters. Its core is the calculation of the relativistic Lorentz force
\begin{align}
 \frac{\text{d}}{\text{d}t} \; \gamma \, \vec{v} = \frac{q}{m c}\left[ \,c \vec{E}(\vec{x},t) \,+ \, \vec{v} \times \vec{B}(\vec{x},t)\, \right], \label{eq:lorentzforce}
 \end{align}
which acts on the particles at position $\vec x$ through the electric $\vec E$ and magnetic fields $\vec B$ of the plasma. The particle velocities are denoted
by $\vec{v}$, $c$ is the speed of light and $\gamma$ represents the Lorentz factor. The \textsc{Gismo--Particles} tool is capable of simulating test particles using
physical masses and charges of electrons and ions. The test particles do not influence the background plasma, i.e. self-consistent backreactions to the plasma are neglected.
A charged particle within a magnetic field will perform a gyromotion with the frequency
\begin{align}
 \Omega = \frac{Z e B}{\gamma m c}
% % \label{eq:gyrofrequenz}
\end{align}
and the Larmor radius
\begin{align}
 r_L = \frac{v}{\Omega}.
\end{align}
A useful numerical approach for solving Eq. (\ref{eq:lorentzforce}) for gyrating particles is the implicit scheme of the Borispush. The basic idea was given by \citetads{boris70}, where the iterations of the Lorentz force are separated in two partial steps. First, the particles are accelerated by the electric field within a half time step. Second, the gyromotion of the particles, which is caused by the magnetic field, is calculated. After that, the electric fields acts again for another half time step to complete the iteration. This approach leads to a discretisation of the Lorentz force with the following set of equations:
\begin{align}
  \vec{v}^{\;t-{\Delta}t/2} & =  \vec{v}^{\;-} - \frac{q \; \vec{E} \; \Delta t^\text{num}}{2 \; m}  \\
   \frac{\vec{v}^{\;+} - \vec{\;v}^{-}}{{\Delta}t} & = \frac{q}{2 \; \gamma m} (\vec{v}^{\;+} + \vec{v}^{\;-}) \times \vec{B} \label{eq:boris2}\\
   \vec{v}^{\;t+{\Delta}t/2} & = \vec{v}^{\;+} + \frac{q \; \vec{E} \; \Delta t^\text{num}}{2 \; m},
\end{align}
where quantities with $\Delta t^\text{num}/2$ denote the discrete half-time steps. To solve this set of equations with respect to $\vec{v}^{\;+}$, the following steps were used:
\begin{align}
  \vec{h}_1 & =  \frac{q \; \vec{B}}{2 \; \gamma^n \; m \; \mathrm{c}} \Delta t^\text{num} \label{eq:borispush1} \\
  \vec{v}' & =  \vec{v}^{\;-} + \vec{v}^{\;-} \times \vec{h}_1 \\
  \vec{h}_2 & =  \frac{2 \; \vec{h}_1}{1 + \vec{h}_1 \cdot \vec{h}_1} \\
  \vec{v}^{\;+} & =  \vec{v}^{\;-} + \vec{v}' \times \vec{h}_2, \label{eq:borispush4}
\end{align}
where the auxiliary vectors $\vec{h}$ were introduced to calculate $\vec{v}^{\;+}$ \citepads{langdon}. By using these equations, the new velocity and location for each particle are then
\begin{align}
 \vec{v}^{\;t+{\Delta}t/2} & = \vec{v}^{\;+} + \frac{q \; \vec{E} \; {\Delta}t}{2 \; m} \\
 \vec{x}^{\;t+{\Delta}t} & =  \vec{x}^{\;t} + \frac{\vec{v}^{\;t+{\Delta}t/2}}{\gamma(\vec{v}^{\;t+{\Delta}t/2})} \; \Delta t^\text{num}.
\end{align}
The advantage of the Borispush is the very high numerical stability. 
The particles are assumed to undergo gyromotions, hence the particle orbits themselves are stable for an arbitrary time discretisation. 
Even in the limit of $\Delta t^\text{num} \gg \Omega^{-1}$, the particle orbit is stable but converges to an adiabatic drift motion. 
The limitation of this method is the correct resolution of the Larmor radius $r_L$. 
If the timestep chosen is  too large, this leads to a big deviation from the analytical $r_L$.

To provide a correct resolution of the gyromotion, \textsc{Gismo--Particles} performs a couple of single-particle gyrations with the used parameters during each start-up phase. 
The Larmor radius is then measured using a three point-circle approximation. If the deviation to the analytical value is above a given accuracy
threshold, the gyrosimulation is performed with a smaller timestep $\Delta t^\text{num}$. This procedure repeats until the accuracy condition is fulfilled. 
Specifically, in our simulations an accuracy of the order of $|r_L-r_\text{measured}|/r_L \approx 10^{-5}$ was used. This ensures that in each simulation the time discretisation is chosen correctly. 
Of course, the discretisation of spatial grid must resolve $r_L$ as well. This means that twice the Larmor radius must be
much larger than a grid cell and smaller than the whole system length.

Ultrarelativistic particle speeds present limitation to the method of the Borispush. In this case the conservation of energy would be violated, since the ideal
ohmic law is not fulfilled anymore. Beyond Lorentz factors of $\gamma \approx 10^3$, fictitious forces start to act and this approach is no longer applicable \citepads{vay08}.

Both parts of \textsc{Gismo} are calculated for each step. After iterating the Els\"asser MHD fields $\vec w^\pm$, they are transformed into the physical,
electric, and magnetic fields which are transferred to \textsc{Gismo--Particles}. Then the Borispush is performed. 
Each particle responds to its local fields, which are calculated by an averaging method via three-dimensional splines
\citepads{2011ASTRA...7...21S,2012ApJ...750..150W}. Periodic spatial boundary
conditions were used; thus the number of particles remained constant in each simulation.

\subsection{Statistical description}\label{sec:statistictransport}

The general statistical approach of particle transport by the relativistic Vlasov equation,
\begin{align}
\pa{ f_T}{ t} + \frac{\vec{p}}{m\gamma} \cdot \pa{ f_T}{ \vec{x}} + \frac{q}{c} \left(c \vec{E}(\vec{x},t)+\vec{v}\times \vec{B}(\vec{x},t)\right) \cdot  \pa{f_T}{\vec{p}} = S_T(\vec{x},\vec{p},t),
\label{eq:relativVlasov}
\end{align}
describes the development of a particle distribution $f_T$ of species $T$ under the influence of the Lorentz force \citepads{schlickibook}. The momentum is denoted by $\vec{p}$ and the term on the RHS $S_T$ is a source term. It is useful to transform this equation into guiding centre coordinates, since we are not interested in the actual position of the gyration, but in the averaged position of the particle. The Vlasov equation then reads 
\begin{align}
 \pa{F_T}{t} + v\mu \pa{F_T}{Z} - \Omega \pa{F_T}{\phi} + \frac{1}{p^2} \pa{}{X_{\sigma}}(\langle p^2 g_{X_{\sigma}} \delta f_T\rangle) = S_T(X_\sigma,t),
 \label{eq:sphaerischeVlasovgemittelt}
\end{align}
where $F_T = \langle f_T \rangle$ is the expectation value of $f_T$ and $X_{\sigma}$ are the guiding centre coordinates with the angle in the perpendicular plane $\phi$ and the pitch-angle cosine $\mu$. These coordinates are given by
\begin{align}
 p_{x}&=p \; \sqrt{1-\mu^{2}}\cos{\phi}\nonumber\\
 p_{y}&=p \; \sqrt{1-\mu^{2}}\sin{\phi}\nonumber\\
 p_{z}&=p \; \mu \nonumber \\
 x&=X-\frac{c}{qB_{0}}p\sqrt{1-\mu^{2}}\sin{\phi}\nonumber\\
%  y&=Y+\frac{c}{qB_{0}}p\sqrt{1-\mu^{2}}\cos{\phi}\nonumber\\
 z&=Z. \label{eq:particlecoords}
\end{align}
The generalised forces $g_{X_{\sigma}}$ represent the effects of the electromagnetic fields and are basically the time derivatives of the denoted variables, e.g. $g_{X_{\sigma}} = \dot X_{\sigma} $.
This equation is in general analytically unsolvable.

A common approach to describing particle transport analytically is the QLT, which was first suggested by \citetads{jokipii66} in the
context of energetic charged particle transport in turbulent magnetic fields. Its core is the assumption of unperturbed particle orbits. This implies that the
fluctuation amplitudes are small, leading to a quasilinear system. The Vlasov equation Eq. (\ref{eq:sphaerischeVlasovgemittelt}) would then simplify to
\begin{align}
 \pa{F_T}{t} + v\mu \pa{F_T}{Z} &- \Omega \pa{F_T}{\phi} = S_T(X_\sigma,t) \nonumber  \\
 &+ \frac{1}{p^2} \pa{}{X_\sigma} \left(  p^2 \pa{F_T}{\hat{X_\sigma}} \underset{D_{X_\sigma \hat{X_\sigma}}}{\underbrace{\int_0^t \text{d}s \langle g_{X_\sigma}g_{\hat{X_\sigma}}(\hat{X_\sigma},s) \rangle}} \right),
\label{eq:final_FokkerPlanck}
\end{align}
where the method of characteristics was applied and the hat symbol represents quantities along the characteristics \citepads{schlickibook}.
This equation is known as the Fokker--Planck equation with the Fokker--Planck coefficients $D_{X_\sigma \hat{X_\sigma}}$. One of the most interesting variables is the pitch-angle diffusion coefficient $D_{\mu\mu}$. It describes the pitch-angle scattering of the particle and is consequently connected to the scattering mean free path, which can be evaluated by the observable angle distribution and Monte Carlo simulations \citepads{2009AguedaVainio}.
Here, scattering means a resonant wave--particle interaction of $n$th order which fulfils the condition
\begin{align}
 k_{\parallel} \,  v_{\parallel} - \omega + n \, \Omega = 0, \quad n \in \mathds{Z}
\label{eq:wave-particle-res}
\end{align}
\citepads{schlickeiser89}, where $\omega$ is the wave frequency and $ k_{\parallel}$ its parallel wavenumber. $\Omega$ is again the particle gyrofrequency
and $v_{\parallel}$ its parallel velocity component. 
Whether a particle with $v_{\parallel}$ interacts resonantly with a wave with $k_{\parallel}$ is determined by the polarisation \citepads{1966PhFl....9.2377K}. 
Because our MHD model consists of pseudo and shear Alfv\'en waves only, certain values of $n$ can be connected to the different types of waves.
The Cherenkov resonance, n=0, is generated only by waves with a compressive magnetic field ($\delta {\bf B} \cdot {\bf B}_0 \neq 0$), 
i.e. pseudo Alfv\'en waves. Purely parallel waves can contribute only to the $n=\pm 1$ resonance, and resonances with $|n|>1$ occur only for waves with nonvanishing perpendicular wavenumber, i.e. with oblique Alfv\'en waves. For a mathematical treatment of the resonant interactions, see Appendix \ref{appendix:derivdmm}.

For further reading and a detailed insight of the derivation of Eq. (\ref{eq:final_FokkerPlanck}) we refer to \citetads{Schlickeiser2002}. 
A serious problem of the QLT is the limited applicability. The approximation of small fluctuations only holds for weak turbulent systems, where
$\delta B/B_0 \ll 1$. This is important for the local field which acts on the invidual particles. For instance, a strong turbulent region would change
effectively the direction of the mean magnetic field and consequently the gyromotion of the particles. Hence the assumption of unperturbed orbits would be
invalid. Another problem of the QLT is the inadequate description of particles propagating perpendicular to the mean magnetic field, $\mu \approx 0$. 
However, regarding Eq. (\ref{eq:wave-particle-res}) a very small parallel particle velocity will generate a singularity for $v_\parallel \rightarrow 0$. 
One aspect of this paper will concentrate on the applicability of the QLT to describe solar energetic particle transport.

\subsection{The Fokker--Planck coefficient $D_{\mu\mu}$}\label{sec:dmm}

We will now focus on the pitch-angle scattering coefficient in more detail. The basic idea is to compare the simulation results of \textsc{Gismo} with the analytical approach.

Our simulations provide us with all the information of the trajectory and velocity of each individual particle within a turbulent plasma. Particle scattering will be considered here in terms of the pitch-angle cosine
\begin{align}
 \mu = \cos(\alpha) = \frac{\vec v \cdot \vec B}{|\vec v ||\vec B|},
\end{align}
which describes the orientation of the particles' velocity vector with respect to the local magnetic field. In the approximation of the QLT, there are different ways to calculate $D_{\mu\mu}$ by using the absolute change of $\mu$ during a certain time interval for a single particle. We used the definition
\begin{align}
D_{\mu\mu} =  \lim_{\Delta t \to \infty} \frac{(\Delta \mu)^2}{2 \, \Delta t} \stackrel{t\gg t_0}{\approx} \frac{(\Delta \mu)^2}{2 \, \Delta t},
\end{align}
where the time interval $\Delta t = t - t_0$ compared to an initial state at $t_0$ is assumed to be large, i.e. the time evolution $t$ has to be sufficient to develop resonant interactions. The necessary time development is discussed in the result section. Instead of using the change of the pitch-angle cosine $\Delta \mu = \mu_e - \mu_0$, the definition with the change of the pitch-angle $\Delta \alpha = \alpha_e - \alpha_0$ itself leads to
\begin{align}
D_{\alpha \alpha} = \frac{(\Delta \alpha)^2}{2 \, \Delta t} = \frac{D_{\mu\mu}}{1-\mu^2},
\label{eq:daa-coeff}
\end{align}
which represents the scattering frequency.

For the analytical approach we use the magnetostatic limit for the calculation of $D_{\mu\mu}$ for pseudo
\begin{align}
D_{\mu\mu,P}\approx & \frac{2\Omega^{2}(1-\mu^{2})}{B_{0}^{2}} \sum_{n\ne0}\int\, d^{3}k\,\pi\delta\left(k_{\parallel}v_{\parallel}+n\Omega\right) \times \nonumber \\
&\left[J_{n}'(v_{\perp}k_{\perp}/\Omega\right]^{2}P_{xx,P}(\mathbf{k})
\label{eq:dmm-sqltP}
\end{align}
and shear Alfv\'en waves
\begin{align}
 D_{\mu\mu,A}=\frac{2\Omega^{2}(1-\mu^{2})}{B_{0}^{2}} &\sum_{n=-\infty}^{\infty}\int\, d^{3}k\,\pi\delta\left(k_{\parallel}v_{\parallel}+n\Omega\right) \times \nonumber \\
 & \left[\frac{n\, J_{n}\left(v_{\perp}k_{\perp}/\Omega\right)}{v_{\perp}k_{\perp}/\Omega}\right]^{2}P_{xx,A}(\mathbf{k}),
 \label{eq:dmm-sqltA}
\end{align}
where $P_{xx}$ is the $xx$-component of the magnetic field fluctuation tensor, which represents the turbulence power spectrum.
These equations are based on \citetads{Schlickeiser2002} and a more detailed derivation is presented in Appendix \ref{appendix:derivdmm}. 
We note that a proper description of the Cherenkov resonance in Eq. 26 is not possible in this formulation. Using $n=0$ yields an undefined mathematical expression.
These coefficients were semi--analytically solved with the set of equations given in Appendix \ref{appendix:discretdmm}. We note that the approach of the magnetostatic QLT (SQLT) assumes the wave frequency to be small compared to the gyro frequency, $\omega \ll \Omega$.

\section{Simulation setup}\label{sec:simsetup}

We focus on the environmental conditions in the solar corona at a distance of three solar radii. The magnetic background field in this region is approximately $B_0 = 0.174 \text{ G}$, which yields, assuming a particle density of $10^5 \text{ cm}^{-3}$, an Alfv\'en speed of $v_A = 1.2 \cdot 10^8 \text{ cm}\text{ s}^{-1}$  \citepads{ramigramm}. This region is of special interest because particle acceleration by CME-driven shocks is strongest there.

The outer length scale of the simulated system is $L_\text{scale}=3.4\cdot 10^8 \text{cm}$. This value is estimated using the growth rate from \citetads{rami-ongenofalfvs}:
\begin{align}
 \Gamma (k) = \frac{\pi \gamma \Omega}{2 n_p v_A} \int \text{d}^3 p \text{ } v\; \mu |k|  \text{ } \delta(|k|-\frac{\Omega}{ v}) f,
\label{eq:growthraterami}
\end{align}
with proton cyclotron frequency $\Omega$, proton speed $v$, the Lorentz factor $\gamma$, the proton number density $n_p$, as well as $\mu$, the pitch-angle cosine, and $f$ the proton distribution function. As the resonance condition, Eq. (\ref{eq:wave-particle-res}) is applied, assuming that the particles are streaming along the field.

A remark on the notation: the wavenumber is in general defined as $k = (2 \pi j)/L$, where $j$ stands for the numerical grid position. For simplicity we used the normalised wavenumbers $k' = k L =(2 \pi \cdot j)$ throughout.

We generated peaks at $k_\parallel'=2\pi\cdot 8$ and $k_\parallel'=2\pi\cdot24$, which can be excited by streaming protons with energies of $E\approx 64 \text{ MeV}$ and $E\approx 7 \text{ MeV}$ respectively. Using the resonance condition, this leads to a length scale of
\begin{align}
 L_\text{scale}= \frac{2 \pi j}{e B_0} \gamma m_p c v \approx 10^8 \text{ cm}.
\end{align}
On this scale the magnetic background field is assumed to be homogeneous.
The spatial resolution is $256^3$ gridpoints, resulting in $128^3$ points in k-space of which $|\vec{k'}|=2 \pi[ 0 \cdots 42 ]$ wave modes are active modes that remain unaffected by (anti)aliasing. The resistivity parameter was set to $\nu = 1$ in numerical units and the hyperdiffusivity coefficient to $h=2$.

The turbulence was simulated using an anisotropic driving mechanism. Energy is, therefore, constantly injected into the simulation volume at the first 5 numerical wavenumbers in perpendicular ($k'_{\perp}=2 \pi [0\cdots4 ] $) and 15 in parallel direction ($k'_{\parallel}=2 \pi [0\cdots14 ] $). 
The reference frame is the solar wind frame.

The anisotropy was chosen for two reasons: first, to mimic the preferential direction of the solar wind, where particles streaming radially away from the Sun
form the Parker spiral. Consequently, these particles can deposit their energy in a parallel direction on different scales. 
This is mainly valid in the vicinity of the Sun, which we are interested in. 
Second, a slab component of solar wind turbulence is also observed at small scales in the parallel direction. 
To ensure turbulence evolution up to high parallel wavenumbers, the driving range was extended along the parallel axis. 
This is necessary because the parallel evolution is much weaker than the perpendicular one \citepads{gsstrong,gsrev}. 
Even though this is primarily a technical aspect to ensure the extent of the spectrum to higher $k_\parallel$, it is
still in line with observations. An isotropic driver would not yield sufficiently turbulent modes at high $k_\parallel$.

The turbulence driving is performed by allocating an amplitude with a phase to the Els\"asser fields within the Fourier space. 
The amplitude follows a power--law of $|\vec{k}|^{-2.5}$ and is initialised using a normal distribution. The phase was randomly chosen between zero and $2\pi$.
These settings are divergence free and Hermitian symmetric. After this initialisation the values were scaled to the desired scenario, which in our case is a
$\delta B/B_0$ ratio of roughly $10^{-2}$. We note that both species, pseudo  and shear Alfv\'en waves, are excited by this type of turbulence driving, but as presented
by \citetads{marongold} the pseudo wave evolution is strongly suppressed.
In this inertial range energy is injected every $0.03s$ (10 MHD steps), which leads to a saturated turbulence, an
equilibrium between dissipation and injection.

A Gaussian-distributed energy peak with purely parallel wavenumber $\vec{k}=k \vec{e_\parallel}$ was injected within the saturated turbulence. 
For this purpose, two different wave modes were chosen. To investigate the physics of an SEP-event, a wavenumber of $k_\parallel=1.5\cdot10^{-7} \text{ cm}^{-1}$ is used. 
This corresponds to a numerical wavenumber of $2\pi \cdot 8$, which is still within the driving range of the turbulence. To represent injection at smaller
scales, a peak is set at $k_\parallel=4.4\cdot10^{-7} \text{ cm}^{-1}$. With  $k'_\parallel= 2\pi \cdot 24$, this wave mode is not in the driving range of the
background turbulence. The injection of energy in these modes was done gradually over a specific time interval.

One important aspect of the calculations of $D_{\mu \mu}$ is the applicability of the QLT. 
The ratio $\delta B/B_0$ is, therefore, the limiting parameter (see Sect. \ref{sec:statistictransport}). 
Consequently, it is of interest to explore peaks at either position under the influence of different magnetic background fields. 
For this purpose another turbulence simulation was performed which uses the same initial conditions, except for a higher mean field $B_0= B_0^h = 1.74$ G and for stability reasons a higher dissipation rate $\nu = 10 \text{ cm}^{2h}\text{s}^{-1}$. 
In the following the smaller field simulation will be denoted by $B_0 = B_0^l = 0.174 \text{ G}$. We note that the $B_0^h$ case is an artificial scenario rather than representative of coronal conditions. Peaks in such high magnetic fields represent SEP energies up to 2.65 GeV. In this case the SEPs cannot generate waves by streaming as the relativistic proton intensities are insufficient.

To summarise, these four simulations with excited wave modes at $k'_\parallel= 2\pi \cdot 8$ and $k'_\parallel= 2\pi \cdot 24$, both within turbulent fields governed by a strong and a weak $B_0$, are the starting point for the test particle simulations and the calculations of the scattering coefficient $D_{\mu \mu}$.

\begin{table*}
\caption{ Parameter setup for the particle simulations.
\label{tab:simparameter}}
\begin{center}
\begin{tabular}{c c c c c c c}\hline \hline
   $B_0$ & $v_A$ & $\nu$ & $k$-grid & $v$ & $\mu_R$& $\mu_R$\\
   $[\text{G}]$ & $[\cms]$ & $[\text{cm}^{2h}\cdot \text{s}^{-1}]$ & & $[\cms]$ & $(k'_\parallel= 2\pi \cdot 8)$& $(k'_\parallel= 2\pi \cdot 24)$\\

\hline
  $0.174$ & $1.2\cdot 10^8$ & $1$ & $128^3$ & $1.21 \cdot 10^{10}$ & $0.86$ & $0.29$\\
\hline
  $0.174$ & $1.2\cdot 10^8$ & $1$ & $256^3$ & $1.21 \cdot 10^{10}$ & $0.86$ & $0.29$\\
\hline
  $1.74$ & $1.2\cdot 10^9$ & $10$ & $128^3$ & $2.9 \cdot 10^{10}$  & $1$ & $0.37$\\
\hline
 \end{tabular}
\tablefoot{The outer length scale was set to $L_\text{scale} = 3.4\cdot 10^8 \text{cm} $ for each simulation. The number density $n_d = 10^5 \text{cm}^{-3}$ connects the background field $B_0$ with the Alfv\'en speed $v_A = B_0 /\sqrt{4 \pi m n_d}$. All of the simulations were performed with $10^5$ test particles with the speed $v$. The $n=1$ resonant $\mu_R$ values for the peaks in each simulation are shown in the two columns on the right.}
\end{center}
\end{table*}

All of the test particle simulations by \textsc{Gismo--Particles} were performed with $10^5$ protons with an initial uniform distribution in $\mu$ and $\phi$ and random positions $\vec x$. This amount has proven to be sufficient in test simulations for good statistics.
These initial conditions aim to provide a complete coverage of the phase space in $\mu$ for the test particles. Thus, we are not interested in the development of special distribution functions.

A constant absolute value of the momentum was chosen so that particles with a certain parallel velocity component fulfil the resonance condition Eq. (\ref{eq:wave-particle-res}). Consequently, a resonant value of $\mu$
\begin{align}
 \mu_R = \frac{\omega-n\,\Omega}{k_\parallel\, v} = \frac{\omega-n\,\Omega}{L_\text{scale}^{-1} \,k'_\parallel\, v}
 \label{eq:mures}
\end{align}
must be within the interval $[-1;1]$ for a given particle speed $v$. Since Eq. (\ref{eq:mures}) is dominated by $\Omega$ and hence $B_0$ for particles propagating significantly faster than the Alfv\'en speed, the particle speed $v$ has to be different for $B_0^l$ and $B_0^h$. The parameter setup and the resonant $\mu_R$ for each simulation is summarised in Table \ref{tab:simparameter}.

To investigate the particle scattering in different stages of the turbulence evolution, multiple simulations were performed. The most promising match between QLT and the simulation results is expected to be in stages with small $\delta B/B_0$. According to this, the particle simulations were performed in the driving phase of the peaks as well as in the decaying phase. A simulation at maximum driven peaks would simply lead to random scattering where no reasonable prediction can be made by QLT.

The simulations with the low magnetic background field $B_0^l$ were also performed with a higher spatial resolution of $512^3$ grid cells, e.g. $256^3$ wave modes. For reasons of clarity the results of these simulations are presented and discussed in the Appendix \ref{appendix:512runs}.

\section{Results and discussion}\label{sec:results}

\subsection{A toy model for wave--particle resonance}
\label{sec:toymodel}
In order to motivate our particle-scattering simulations, we want to present a simple model. Because of the limited applicability of QLT, it is necessary to understand its range of validity. In detail, the $\delta B/B_0$ ratio and the time development are of utmost importance. Different simplified scenarios are therefore presented below.

We consider a circularly polarised Alfv\'en wave with \mbox{$\vec k = (0,0,k_\parallel)^T$}. Its magnetic field is described by
\begin{align}
 \vec B = \vec B_0 + \delta B \left( \vec e_x \cos(\pm k z + \psi) + \vec e_y \sin(\pm k z + \psi) \right ), \label{eq:alfvenwavemagneticfield}
\end{align}
in the wave frame where $B_0 \gg \delta B$ is assumed. The interaction between particles and the wave will then change the particles' pitch-angles via scatterings. A detailed description of this process as well as a derivation for the time dependency of $\mu$ is given in Appendix \ref{appendix:deltamu}. Under the assumption given by Eq. (\ref{eq:alfvenwavemagneticfield}). the expression of the time evolution of $\mu$ in Eq. (\ref{eq:mustreuung}) will simplify to
\begin{align}
 \Delta \mu^\pm (t,\psi) = \Omega \, &\sqrt{(1-\mu^2)} \, \frac{\delta B}{B_0} \times \nonumber \\
 &\frac{\cos \psi - \cos\left[(\pm k v \mu - \Omega)\Delta t + \psi \right]}{\pm k v \mu - \Omega}, \label{eq:wtr-mu}
\end{align}
when the calculation is performed along the unperturbed orbit, $\mu=$ constant. The change in $\mu$ is then maximal at the phase
\begin{align}
 \psi^\pm_M = \arctan \left[ \frac{\sin(\pm k v \mu - \Omega)\Delta t}{1-\cos(\pm k v \mu - \Omega)\Delta t} \right]
 \label{eq:wtr-psi}
\end{align}
and its identical solutions at periods $\psi^\pm_M + n \pi$.

To study this model, a single linearly polarised Alfv\'en wave was simulated, propagating undisturbed with $v_A=10^7 \cms$ towards positive $z$-direction ($\vec w^+$) with a purely parallel wave vector $\vec k' = 2\pi \cdot (0,0,1)^T$. It should be noted that the magnetic field given by
Eq. (\ref{eq:alfvenwavemagneticfield}) does not describe a propagating wave but a static disturbance. 
Linear polarisation was chosen to cover both resonances at $\pm \mu_R$ (respectively both signs in Eq. (\ref{eq:wtr-mu})), because interactions with $\mu<0$ are caused by left-handed circular polarised waves and $\mu>0$ by right-handed ones. 
The wave amplitude was set to $\delta B/B_0=0.1$, where $B_0=4.34 \cdot 10^{-4}$ G. 
The outer length scale of the simulation cube was set to $10^8$ cm.

\begin{figure}[ht]
  \begin{center}
    \mbox{
          \includegraphics[width=1. \columnwidth]{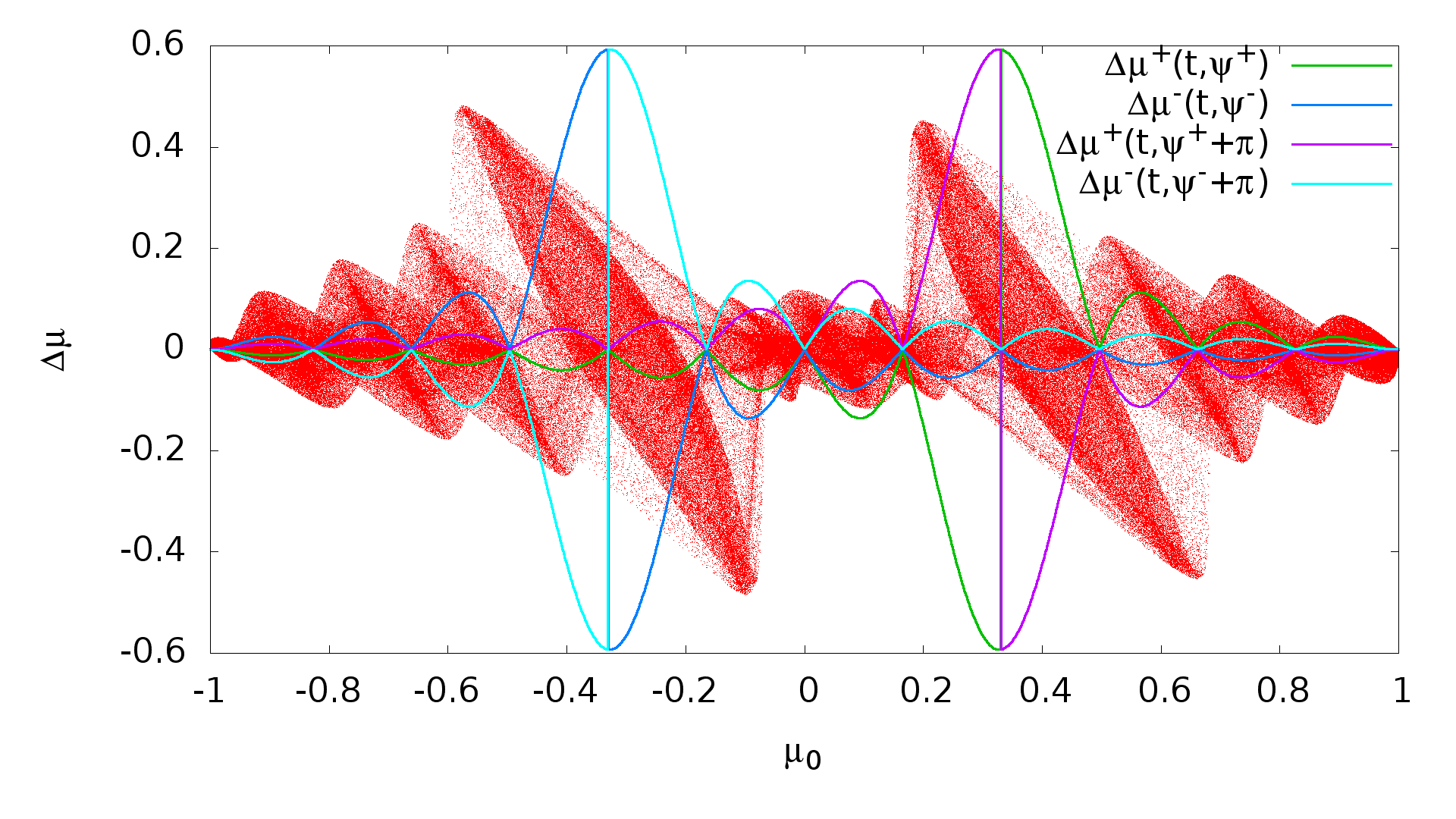}
         }
    \caption{Result of a toy model wave--particle resonance after two gyration periods with a wave amplitude of  $\delta B/B_0=0.1$. The pitch-angle scattering is presented in terms of change in $\Delta \mu$ dependent on the particles intial $\mu_0$. Each dot represents an individual particle. The theoretical, quasilinear prediction of the maximal change in $\mu$ by Eq. (\ref{eq:wtr-mu}) deviates clearly from the observed scattering. This type of plot is hereafter denoted as scatter plot.}
\label{fig:wtr-2gyr-vielamp}
  \end{center}
\end{figure}

In \textsc{Gismo--Particles}, $5 \cdot 10^5$ protons with random $\mu$, $\phi$ and position $\vec x$ were initialised. The absolute value of $\vec{v}$ was set to $2\cdot10^8 \cms$, where particles for $n=-1$ and  $n=+1$  are resonant at $\mu_{R1}=-0.28$ and $\mu_{R2}=+0.38$ according to Eq. (\ref{eq:wave-particle-res}).

In Fig. \ref{fig:wtr-2gyr-vielamp}, we show the pitch-angle change of the simulated particles and the QLT-predicted change as a function of the initial pitch-angle for wave amplitude $\delta B/B=0.1$ after two gyration periods. 
The simulated particles are represented as a scatter plot, whereas the QLT result for different wave modes are represented by curves. As expected, two maxima develop at the predicted positions $\mu_{R1}$ and $\mu_{R2}$.
The adjoining maxima are caused by particles interacting with the wave nonresonantly, henceforth named nonresonant ballistic interactions. 
These decrease their amplitudes with increasing time. 
Also, their positions in $\mu$ do not remain constant, like the resonant interactions, because the
ballistic maxima become more numerous and move closer to the true resonance. The tilt of the maxima deviates from the prediction of Eq. (\ref{eq:wtr-mu}), as clearly shown by comparison with the theoretical curves. 
Furthermore, the resonances of Eq. (\ref{eq:wtr-mu}) are located symmetrically at $\mu_r=\pm0.33$.
This shift can be explained by the choice of the inertial system. The derived equation is within the wave frame. 
The transformation into the laboratory frame (under the assumption of Galilean transformation) via
\begin{align}
 \mu' \approx  \mu - \frac{v_A}{v}
\end{align}
will fit to the observed position in $\mu$. 
This shift is caused by the magnetostatic approximation, since the difference between Eq. \ref{eq:mures} and its magnetostatic version ($\omega=0$) is exactly the term $v_A/v$. If this ratio is small, as in most of our simulations, the effect is negligible. However, for the simulations with stronger $B_0^h$ this poses a problem, which will be discussed in the results section.

The tilt cannot be explained by Eq. (\ref{eq:wtr-mu}) due to its dependence on the QLT. 
It stems from the distortion of the orbit by the wave's $\delta B$ and consequently, arises by the presentation of $\Delta \mu$ as a function of the initial pitch-angle, $\mu_0$. 
Because of the finite $\delta B/B_0$--ratio, the assumption of unperturbed orbits, which is the basic approximation in QLT, does not hold anymore. 
The change of $\mu$ should therefore be calculated as an average over the whole trajectory of each particle, which would introduce a nonlinear correction to the theory. 
A simple solution to correct the tilt is the presentation of $\Delta \mu$
dependent not on the intial $\mu_0$, but on the mean value $\langle \mu \rangle$ of the initial and the final value of the pitch-angle cosine. 
We note that the angle of the tilt does not depend on $\delta B/B_0$, but the scattering amplitude does. 
Consequently, the tilt is more visible and influences further calculations more as the 
$\delta B/B_0$--ratio increases. This visibility is called \emph{effective tilt} hereafter.

\begin{figure}[ht]
  \begin{center}
    \mbox{
          \includegraphics[width=1. \columnwidth]{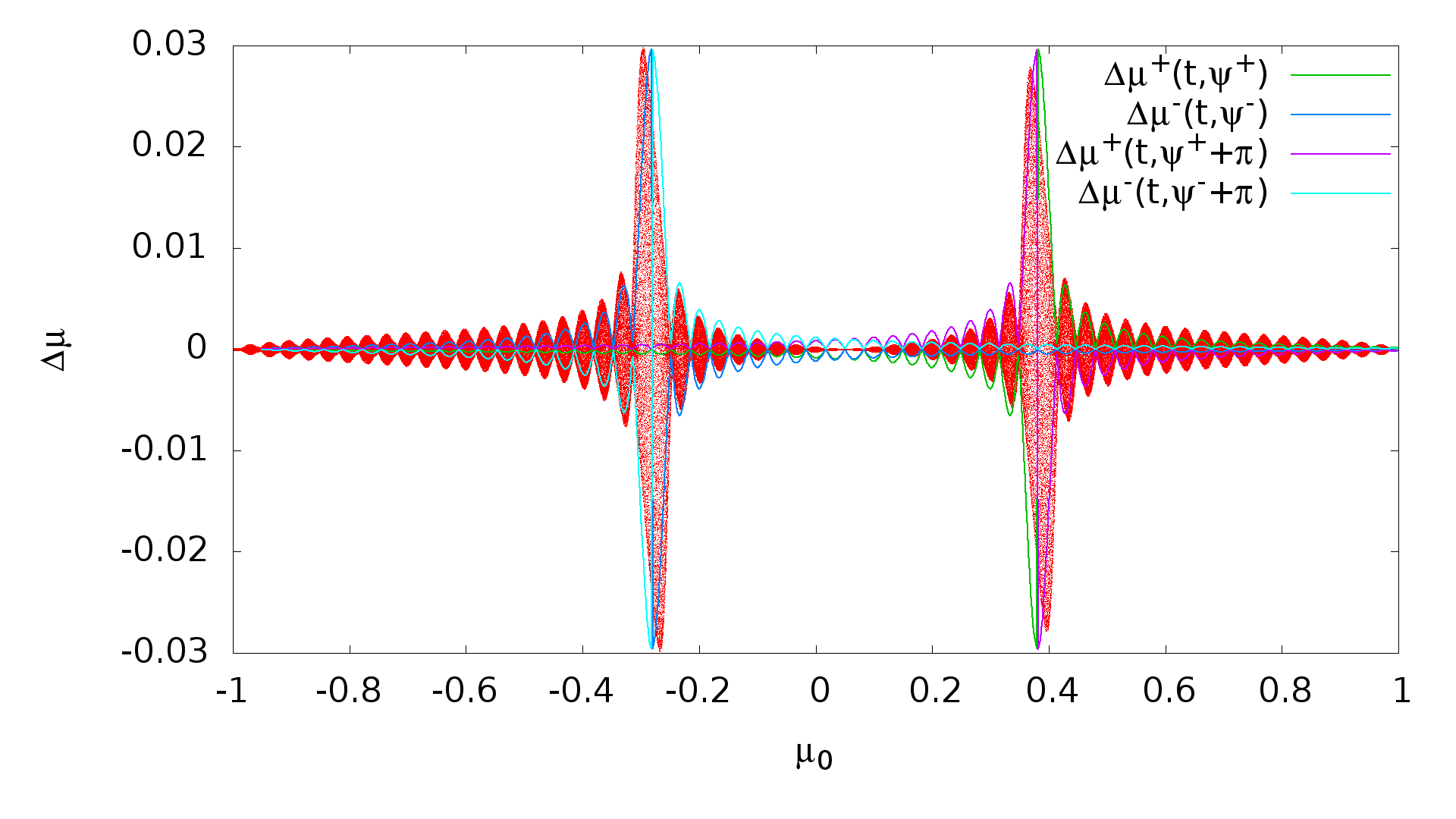}
         }
    \caption{Result of a toy model wave--particle resonance after ten gyration periods with a wave amplitude of  $\delta B/B_0=0.001$ . Each dot represents an individual particle scattered from its initial $\mu_0$ by $\Delta \mu$. The prediction by Eq. (\ref{eq:wtr-mu}) has been transformed into the lab frame and now describes the observed scattering very well.}
\label{fig:wtr-10gyr-wenigamp}
  \end{center}
\end{figure}

In order to prove the explanations given above, we performed a similar simulation with smaller wave amplitude and longer time development. As shown in Fig. \ref{fig:wtr-10gyr-wenigamp} the QLT is valid within this scenario. As the scattering amplitude is significantly smaller, the effect of the tilt resulting from using the initial pitch-angle is negligible and the time evolution leads to a clear resonance as the side maxima become smaller. But even under these convenient conditions the deviation from the QLT is still visible.

\begin{figure}[ht]
  \begin{center}
    \mbox{
          \includegraphics[width=1. \columnwidth]{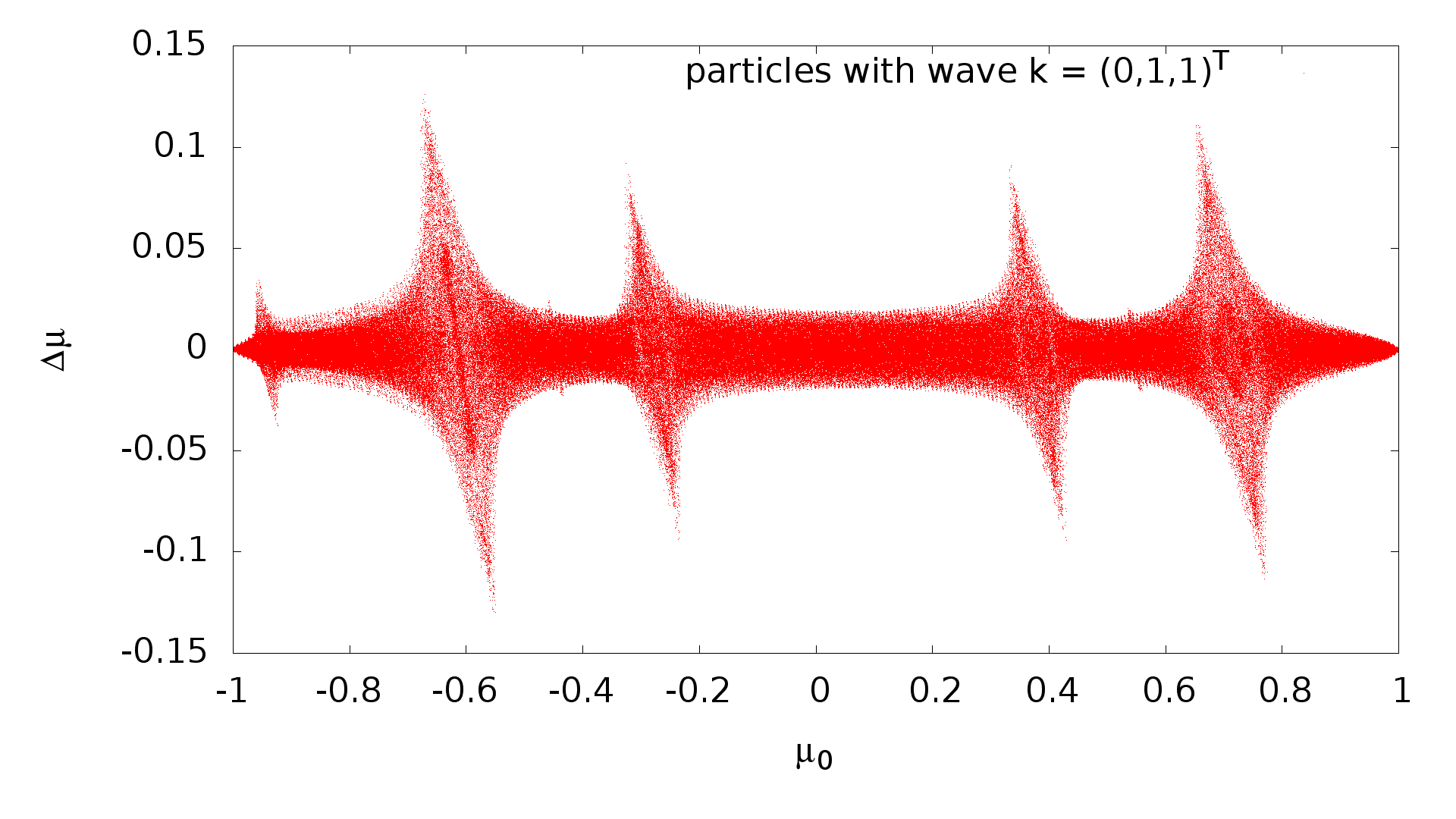}
         }
    \caption{Result of a toy model wave--particle resonance after 50 gyration periods with an oblique wave with an amplitude of  $\delta B/B_0=0.001$. As expected, also higher orders of resonances with $n=[-3,-2,-1,1,2]$ also developed. At this timestep Eq. (\ref{eq:wtr-mu}) would  again deviate significantly, because only few particles would fulfil the resonance condition, where others are still disturbed by the finite distortion of $B_0$ and thus interact nonresonantly, which produces the finite ballistic background.}
\label{fig:wtr-50gyr-wenigamp}
  \end{center}
\end{figure}

The last case of our toy model uses an oblique wave with $\vec k' = 2\pi \cdot(0,1,1)^T$, which was simulated under the same conditions as before. 
The result in Fig. \ref{fig:wtr-50gyr-wenigamp} clearly shows the higher orders of resonance with $n=[-3,-2,-1,1,2]$. A prediction of Eq. (\ref{eq:wtr-mu}) is
not applied, since it would not describe the results of the simulation very well for two reasons. First, the assumption that the prediction by QLT becomes
better with $t\rightarrow \infty$ would not hold because of the finite $\delta B/B_0$ and hence perturbed orbits. 
Second, the nonvanishing Bessel functions have to be taken into account and would modify Eq. (\ref{eq:wtr-mu}). 
But although the ballistic maxima become significant in number and generate a nonresonant
background scattering, the resonances are clearly visible. 
Another interesting fact is that the scattering caused by resonance at $|n|=2$ exceeds the ones at $|n|=1$. 
This means that in the case of an oblique wave, higher order resonances can become even more important than the fundamental ones.
The time development for the purely parallel wave mode at 50 gyration periods showed expectedly only the $|n|=1$ resonances (not shown in Fig. \ref{fig:wtr-50gyr-wenigamp} for reasons of clarity).

\subsection{Results of particle scattering in turbulence with amplified modes}\label{sec:particlesims}

\begin{figure}[ht]
  \begin{center}
    \mbox{
          \includegraphics[width=1. \columnwidth]{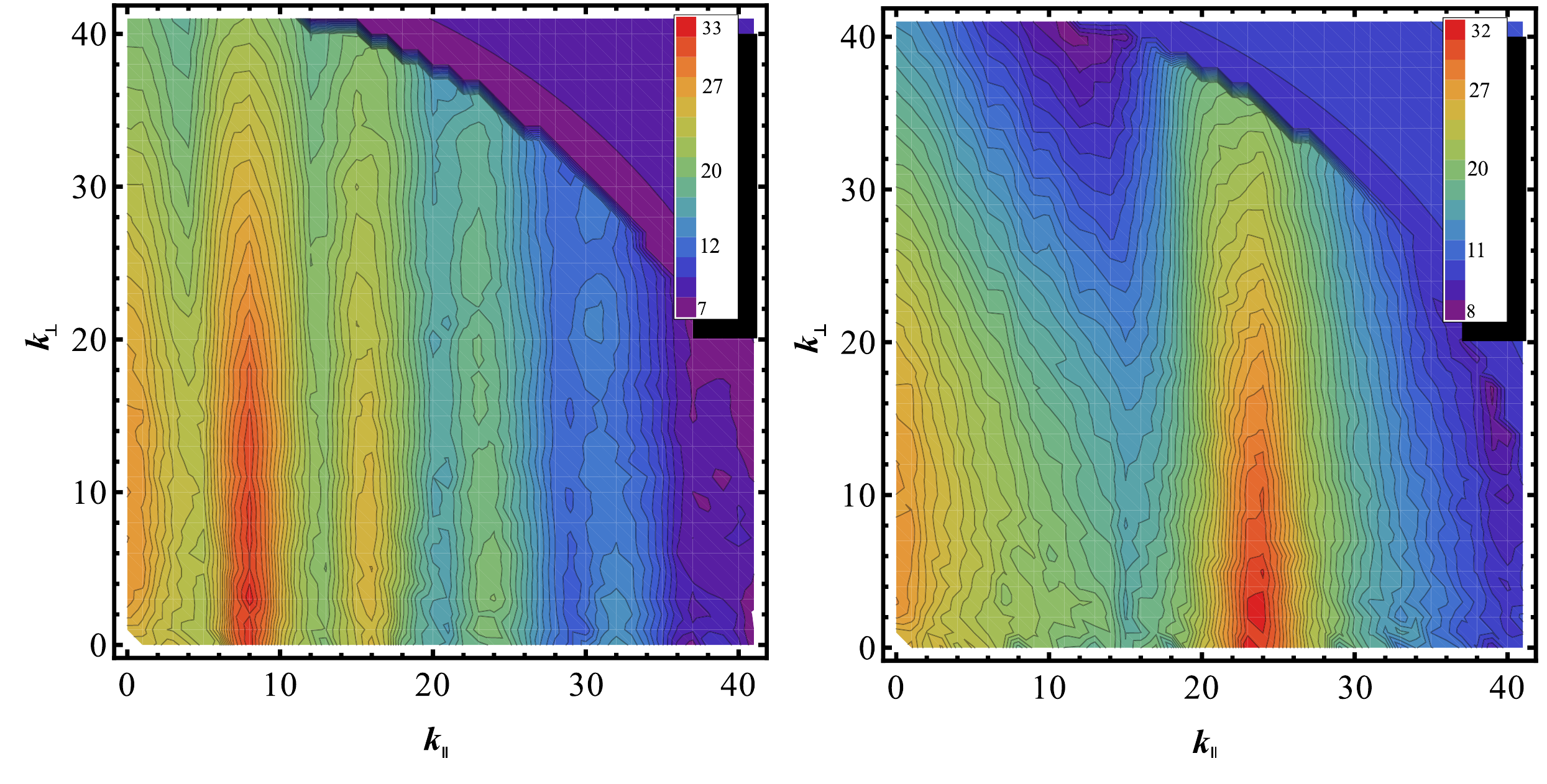}
         }
    \caption{Two-dimensional magnetic energy spectra of both peaks in the simulation with $B_0^l=0.174$ G. These are the base scenarios for the particle simulations during the decay stage of the peaks. Therefore the left figure shows the state for the $k'_\parallel = 2 \pi \cdot 8$ peak at $t=17$ s. The right figure shows $k'_\parallel = 2 \pi \cdot 24$ at $t=5.1$ s. The colours indicate the logarithm of the total spectral energy.}
\label{fig:v31-spherplots}
  \end{center}
\end{figure}

We first present selected results from \citetads{lange2012}, which are the base of our particle simulations. However, we do not discuss the evolution of the peaks in detail, but focus on the stages of the driving and decay of the peaks on which our particle simulations were performed.

\begin{figure}[ht]
  \begin{center}
    \mbox{
          \includegraphics[width=1. \columnwidth]{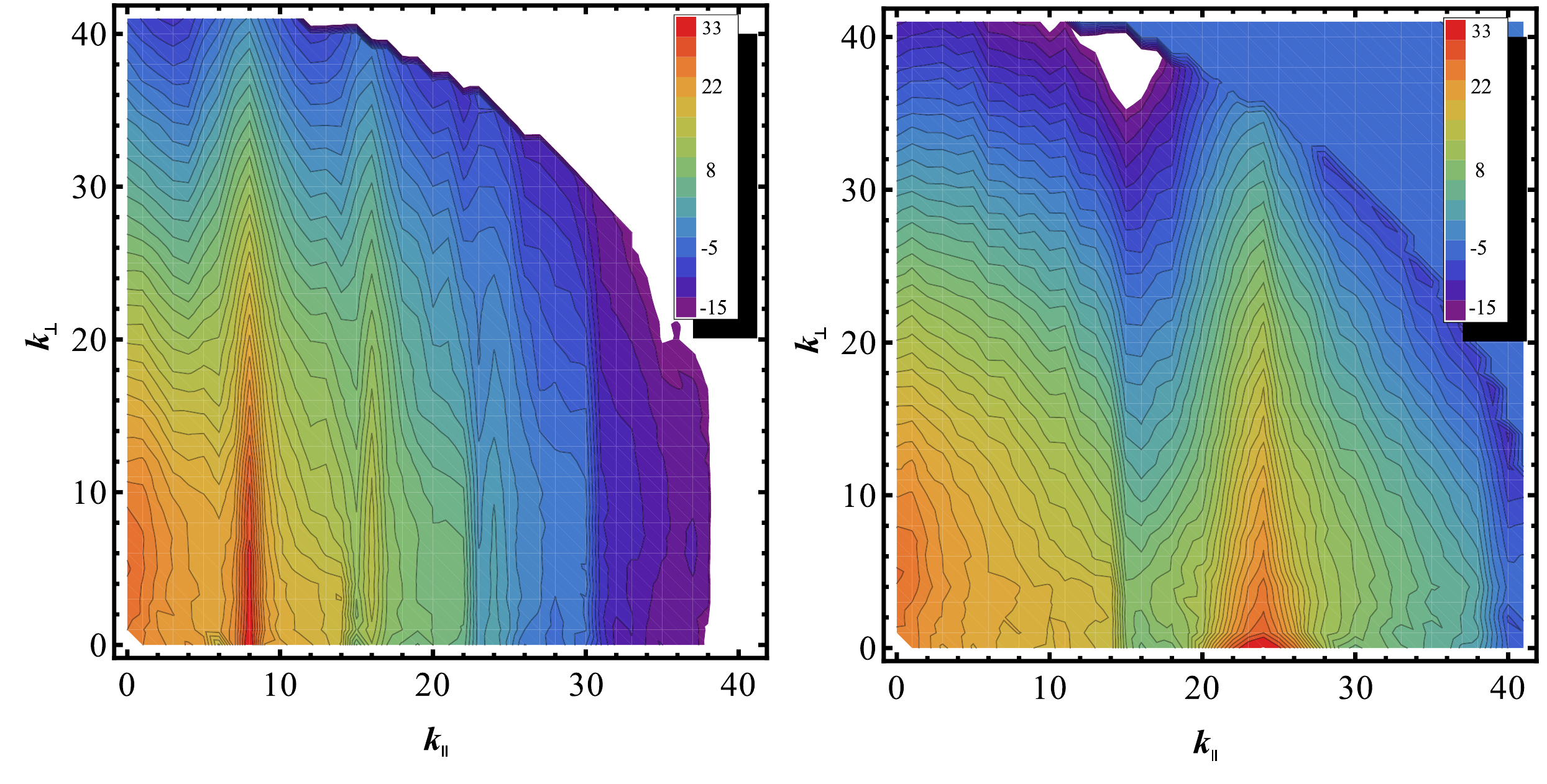}
         }
    \caption{Two-dimensional magnetic energy spectra of both peaks in the simulation with $B_0^h=1.74$ G. The decay stage of the peaks is presented. The left figure shows the state for the peak at $k'_\parallel = 2 \pi \cdot 8$ at $t=13.6$ s, whereas the right one shows $k'_\parallel = 2 \pi \cdot 24$ at $t=2.04$ s. The colours indicate the logarithm of the total spectral energy.}
\label{fig:v35-spherplots}
  \end{center}
\end{figure}

In Figs. \ref{fig:v31-spherplots} and \ref{fig:v35-spherplots} we show the situation for the decaying peaks. The timesteps of these figures are the starting points of our particle simulations for the decay stage. We note that the timesteps were chosen to retrieve roughly the same order of magnitude in $\delta B/B_0$ at the peak positions, which are $10^{-3}$. All of the peaks show a strong perpendicular evolution at this stage. Furthermore, higher harmonics of $k'_\parallel = 2 \pi \cdot 8$  at positions 16, 24, 32, 40 are  present. These harmonics, however, do not interact dominantly with the particles as they have already lost most of their energy compared to the maximum driven stage (not shown here,) as indicated by the logarithmic colour scaling.

\begin{figure}[ht]
  \begin{center}
    \mbox{
          \includegraphics[width=1. \columnwidth]{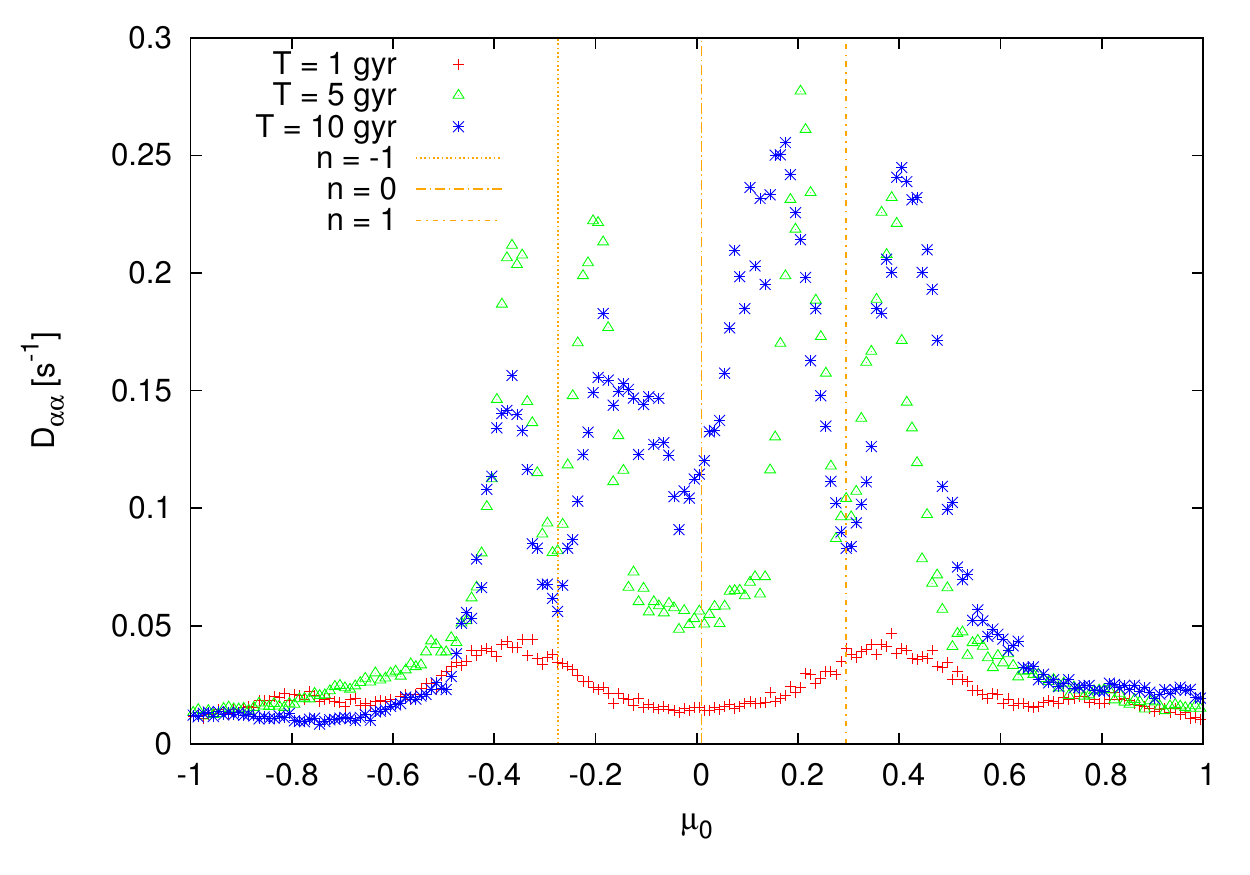}
         }
    \caption{Time evolution of the pitch-angle scattering coefficient $D_{\alpha \alpha}$. The setup used is the low $B_0^l, \; 256^3$ gridsize, peak position $k'_\parallel = 2 \pi \cdot 24$, driven stage. The vertical lines represent the predicted positions of the resonances according to Eq. (\ref{eq:mures}).}
\label{fig:v31-small24-driven-Daa}
  \end{center}
\end{figure}

As shown in the previous section on the toy model, the resonance takes several gyration periods to develop. On the other hand, if the particles are simulated
for too long, the interactions with the turbulence will smooth the resonances and lead to unstructured scattering. 
This is again caused by the finite perturbation of the particle orbits, which are assumed to be negligible for the applicability of QLT. The resonances would not converge to a $\delta$-function as predicted by QLT, but be scattered ballistically at the finite distortions of the fields.
We find that a reasonable timescale is between 10 and 30 gyration periods for parameters used in this paper. 
For weaker turbulence, even results at 50 gyrations might yield a structure, 
but in most cases the resonances develop clearly in shorter timescales. The pitch-angle scattering coefficient $D_{\alpha \alpha}$ was calculated for different
stages of the evolution. In Fig. \ref{fig:v31-small24-driven-Daa} we show $D_{\alpha \alpha}$ for the first simulation setup with $B_0^l = 0.174$ G within the
$256^3$ grid, with the peak position $k'_\parallel = 2 \pi \cdot 24$ at the driven stage.  We observe a convergence between 5 and 10 gyration periods. 
As presented in Table \ref{tab:simparameter}, the resonant interaction with the $k'_\parallel = 2 \pi \cdot 24$ mode is located at $\mu_R=0.29$,
 whereas Fig.  \ref{fig:v31-small24-driven-Daa} shows two maxima at $\mu=0.2$ and $\mu=0.4$, which seem to move away from each other during the time development.

\begin{figure}[ht]
  \begin{center}
    \mbox{
          \includegraphics[width=1. \columnwidth]{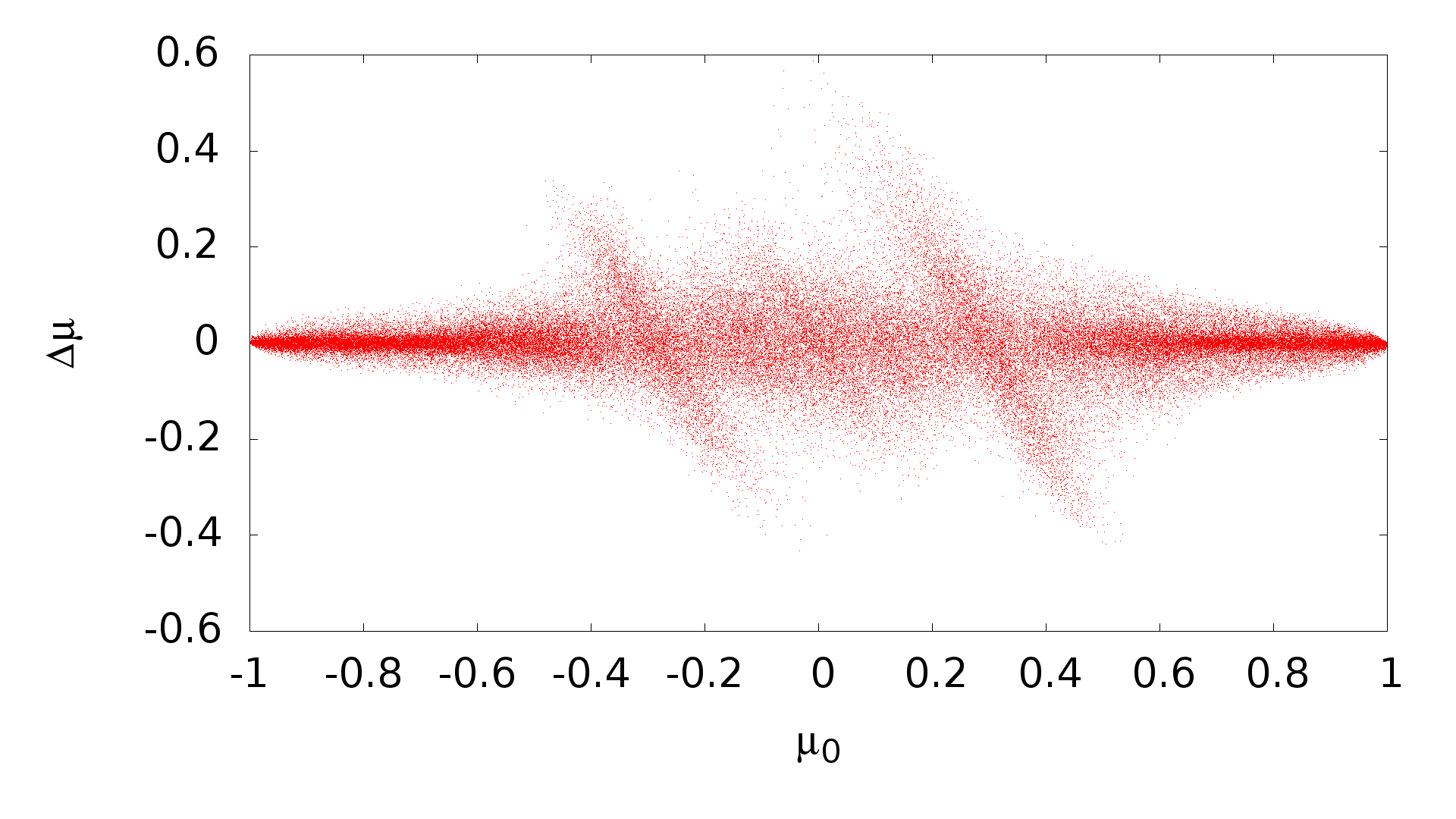}
         }
    \caption{Scatter plot for the low $B_0^l, \; 256^3$ gridsize, peak position $k'_\parallel = 2 \pi \cdot 24$, driven stage, $t=10$ gyration periods. Each dot represents the total change of $\mu$ of an individual particle. Three resonance patterns are visible and centred at $\mu_R=-0.27, 0, 0.29$. }
\label{fig:v31-small24-driven-deltamue}
  \end{center}
\end{figure}

The real resonant structure is revealed by the scatter plot in Fig. \ref{fig:v31-small24-driven-deltamue}. Indeed, the resonance is centred at the predicted
position, but tilted for the same reasons presented with the toy model. The tilt spreads the particles to a wide $\Delta\mu$ range, which results in
splitting of the maximum of the pitch-angle diffusion coefficient into two maxima at both sides of the resonant pitch-angle $\mu_R=0.29$. This is because the
calculation of $D_{\alpha \alpha}$ in the QLT with Eq. (\ref{eq:daa-coeff}) is not dependent on the sign of $\Delta \mu$. 
Consequently, the scattering coefficient is mapped due to the square value of $\Delta \mu$ to two different maxima. 
This also explains the movement of the maxima between five and ten gyration periods in Fig. \ref{fig:v31-small24-driven-Daa}, because $\Delta \mu$ increases with time. 
The smaller resonance at $n=-1$, i.e. $\mu_R=-0.27$, is caused by the different polarisations of the peaked mode. 
That means resonances with $\mu<0$ are caused by left-handed circular polarised parts of the peaked mode and $\mu>0$ by right-handed ones. 
Furthermore, in the scatter plot the Cherenkov resonance $n=0$ is visible. 
It is hardly observable in the  $D_{\alpha \alpha}$ plot due to the dominant $|n|=1$ resonances and their tilt.

\begin{figure}[ht]
  \begin{center}
    \mbox{
          \includegraphics[width=1. \columnwidth]{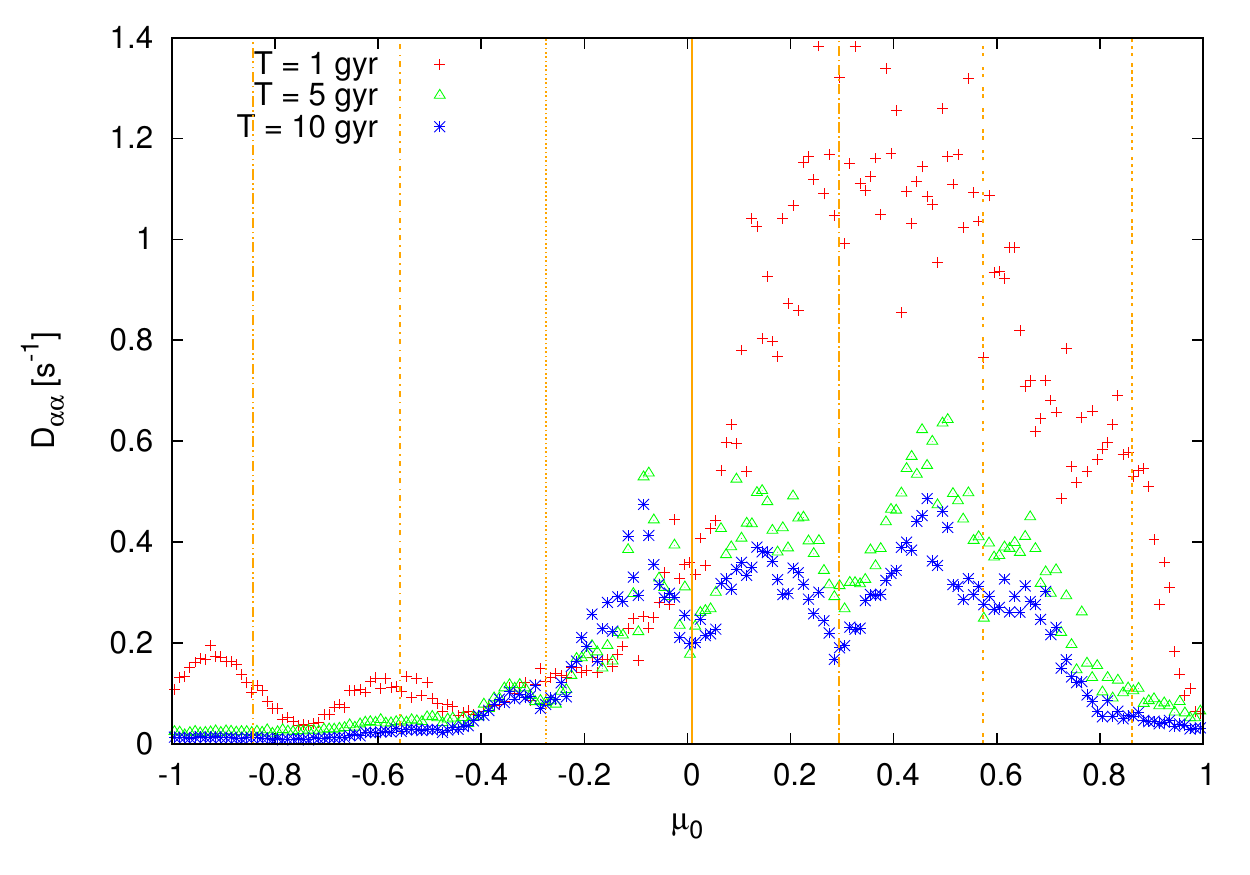}
         }
    \caption{Time evolution of the pitch-angle scattering coefficient $D_{\alpha \alpha}$. The setup used is the low $B_0^l, \; 256^3$ gridsize, peak position $k'_\parallel = 2 \pi \cdot 24$, decay stage. The structure became more complex due to higher order resonances. The vertical lines represent the predicted positions of the resonances according to Eq. (\ref{eq:mures}).}
\label{fig:v31-small24-decay-Daa}
  \end{center}
\end{figure}

\begin{figure}[ht]
  \begin{center}
    \mbox{
          \includegraphics[width=1. \columnwidth]{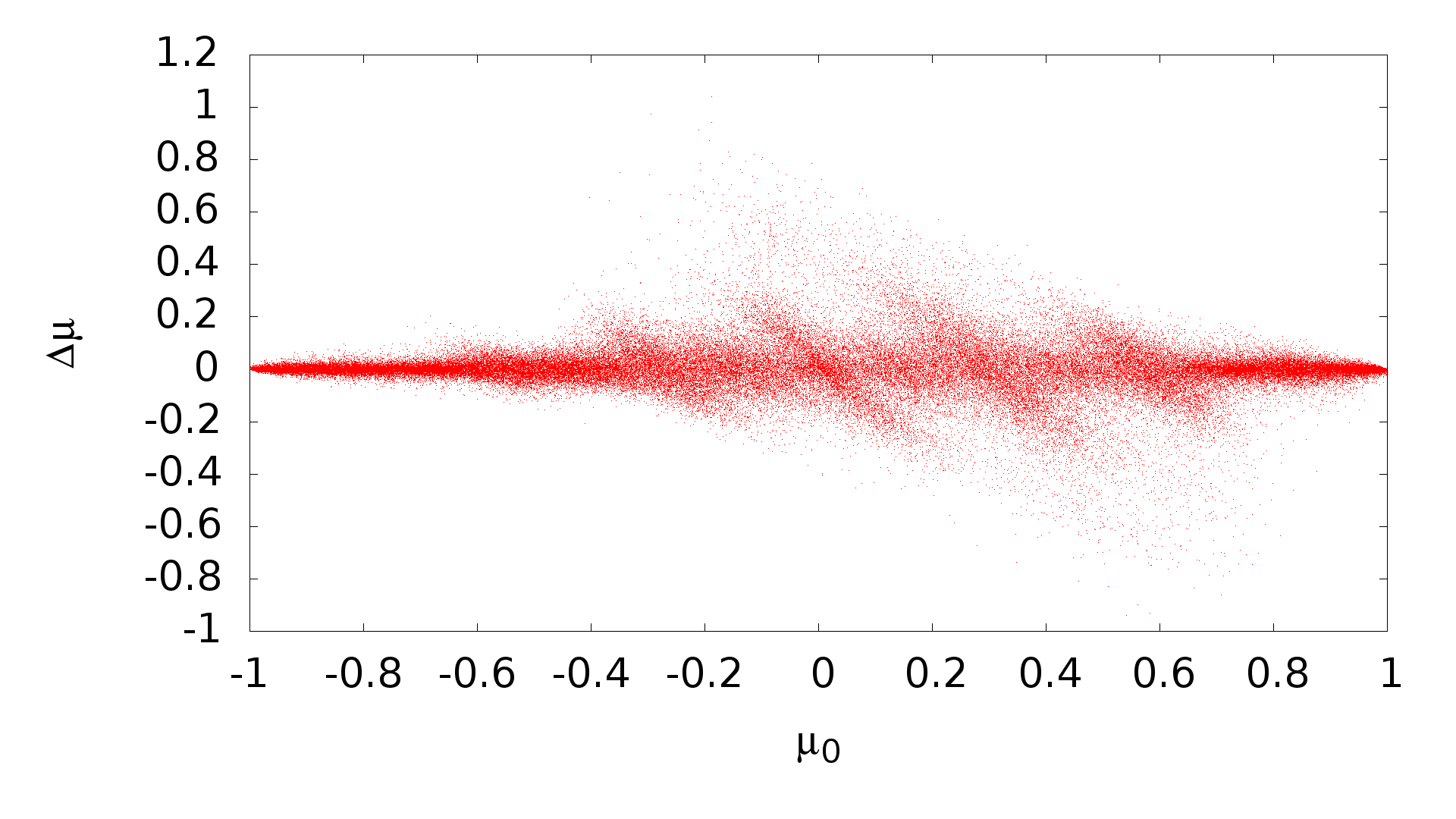}
         }
    \caption{Scatter plot for the low $B_0^l, \; 256^3$ gridsize, peak position $k'_\parallel = 2 \pi \cdot 24$, decay stage, $t=10$ gyration periods. Several resonances occurred during the decay stage of the peak. The scatter plot reveals the resonances more accurately than the $D_{\alpha \alpha}$ shown in Fig. \ref{fig:v31-small24-decay-Daa}; especially smaller substructures are visible.}
\label{fig:v31-small24-decay-deltamue}
  \end{center}
\end{figure}

The test particle simulation with the decay phase of the peaked mode at $k'_\parallel = 2 \pi \cdot 24$ shows resonant interactions beyond the fundamental resonance. For example, the maximum located near $\mu=0.6$ in Fig. \ref{fig:v31-small24-decay-Daa} represents the $n=2$ interaction. We note that the tilt in the corresponding scatter plot in Fig. \ref{fig:v31-small24-decay-deltamue} causes again the split of the maximum in $D_{\alpha\alpha}$.

As already shown by the toy model, these higher orders are generated by oblique Alfv\'en waves. 
Those modes clearly exist in the decay stage of the peaked mode. 
For example, in Fig. \ref{fig:v31-spherplots} the maximum of the turbulence energy spectrum is shifted from the $k_\parallel$ axis towards higher perpendicular wave modes. 
However, within the driven stage of the peak the energy is injected in purely parallel modes, which causes the dominant $|n|=1$ resonances. 
Furthermore, the resonance pattern in Fig. \ref{fig:v31-small24-decay-deltamue} indicates that the left-handed circular polarised parts of the peaked mode decayed faster, which lowers the scattering frequency for particles with $\mu<0$.
While this observation needs further investigation, it seems to be connected to the turbulence evolution, since the wave-particle toymodel does not show this behaviour. Additionally, the magnetic background field plays a role, which is discussed for the results of the other parameter setup below.
Again the Cherenkov resonance $n=0$ is visible.

\begin{figure}[ht]
  \begin{center}
    \mbox{
          \includegraphics[width=1. \columnwidth]{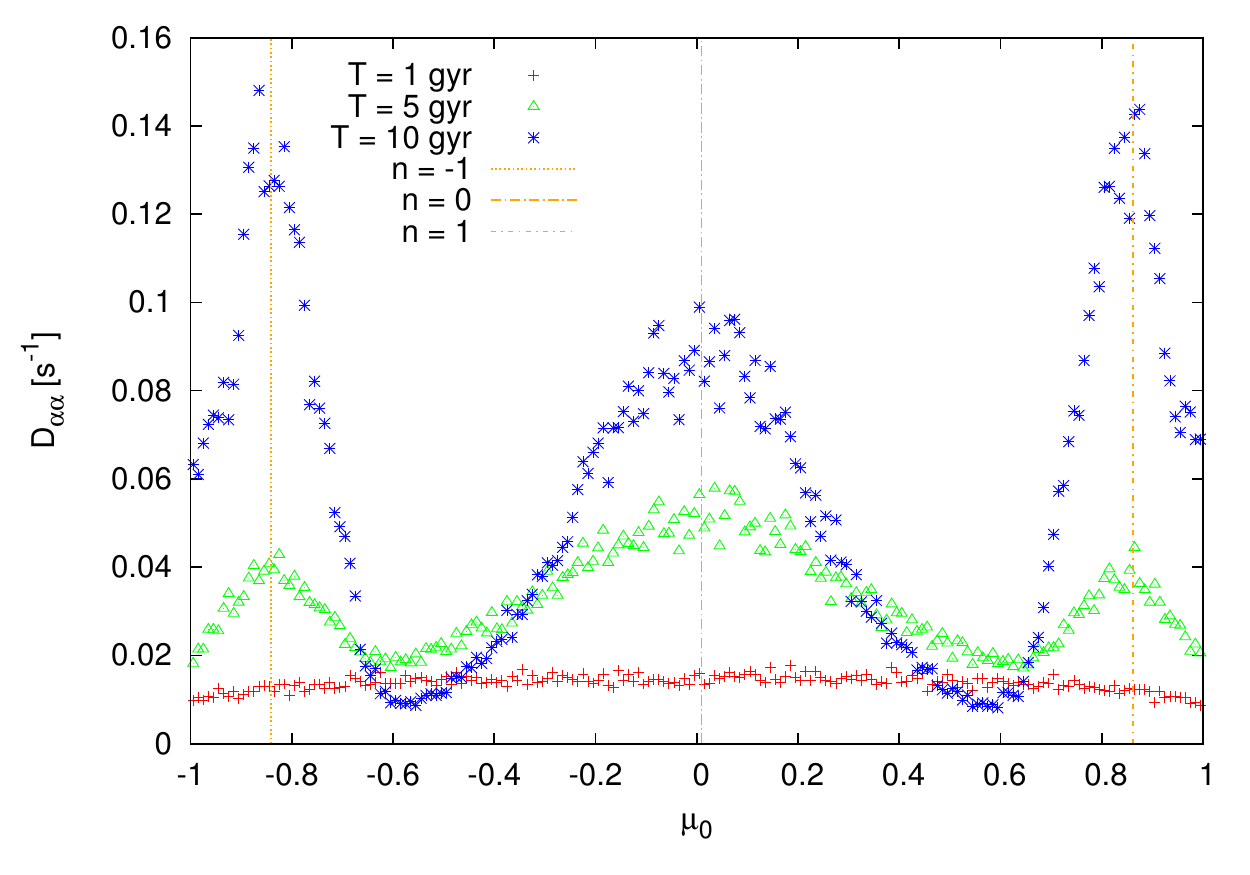}
         }
    \caption{Time evolution of the pitch-angle scattering coefficient $D_{\alpha \alpha}$. The  setup used is the low $B_0^l, \; 256^3$ gridsize, peak position $k'_\parallel = 2 \pi \cdot 8$, driven stage. The resonances occurred at their predicted position.}
\label{fig:v31-small8-driven-Daa}
  \end{center}
\end{figure}

The results for the particle scattering at the $k'_\parallel = 2 \pi \cdot 8$ driven peak are presented in Fig. \ref{fig:v31-small8-driven-Daa}. During the driven stage of the peak, the test particles interacted resonantly at $|n|=1$ and the predicted pitch-angle $\mu_R=0.86$. The effect of the tilt on the resonances is smaller compared to the $k'_\parallel = 2 \pi \cdot 24$ peak. Consequently, the maxima in the scattering coefficient are not split.

\begin{figure}[ht]
  \begin{center}
    \mbox{
          \includegraphics[width=1. \columnwidth]{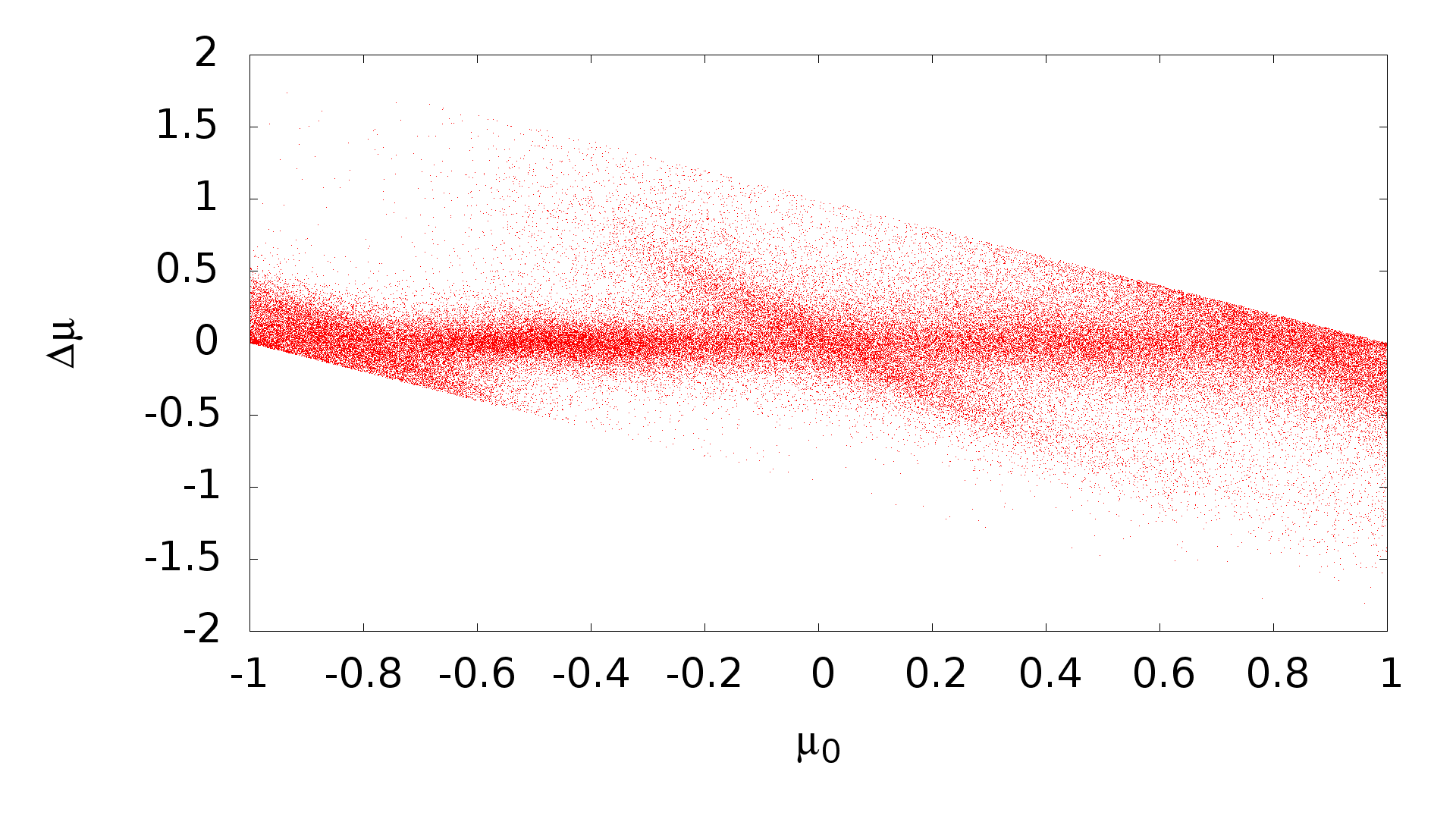}
         }
    \caption{Scatter plot for the low $B_0^l, \; 256^3$ gridsize, peak position $k'_\parallel = 2 \pi \cdot 8$, decay stage, $t=10$ gyration periods. The strong scattering led to strongly tilted structures. Several particles reach the maximum change of $\mu$ as indicated by the sharp straight lines.}
\label{fig:v31-small8-decay-deltamue}
  \end{center}
\end{figure}

The situation is different for the decay stage. The energy has spread significantly to oblique modes (see Fig. \ref{fig:v31-spherplots}, left-hand frame). This leads to stronger scattering and hence a stronger tilt, which is clearly visible in Fig. \ref{fig:v31-small8-decay-deltamue}. Even the sharp lines of the maximum change of $\mu$ are observed, because several particles were strongly scattered. In this state, $D_{\alpha \alpha}$ has no clear structure.

\begin{figure}[ht]
  \begin{center}
    \mbox{
          \includegraphics[width=1. \columnwidth]{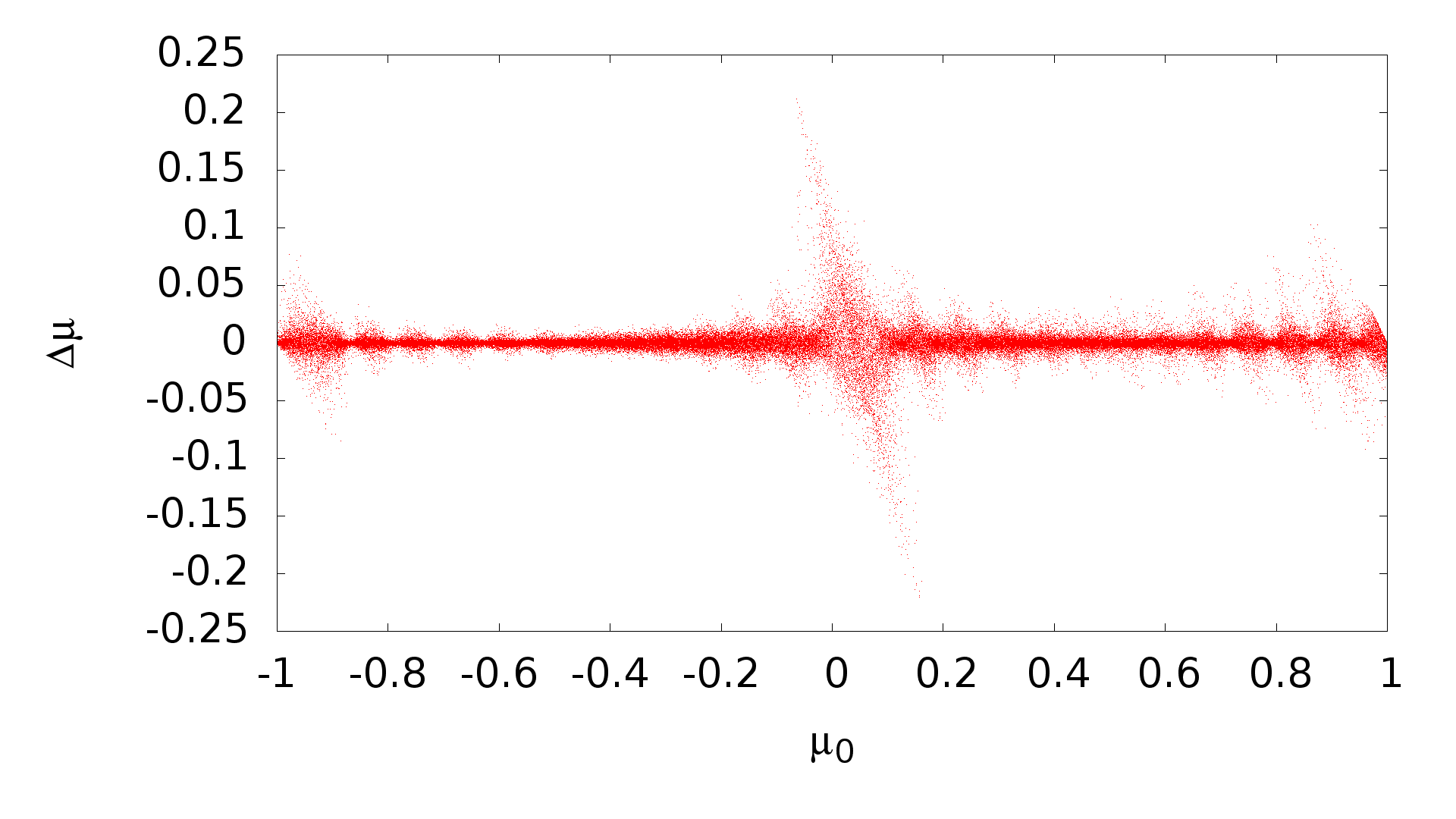}
         }
    \caption{Scatter plots for the high $B_0^h, \; 256^3$ gridsize, peak position $k'_\parallel = 2 \pi \cdot 8$, decay stage, $t=50$ gyration periods. As in the toy model, the reduced $\delta B/B_0$ ratio leads to decreased effective tilts of the resonances. After a long simulation time of 50 gyrations, the resonant interactions become very narrow. The small increase of the scattering near $\mu_0=0.5$ might be caused by the $k'_\parallel = 2 \pi \cdot 16$ higher harmonic.}
\label{fig:v35-small8-decay-deltamue-50gyr}
  \end{center}
\end{figure}

\begin{figure}[ht]
  \begin{center}
    \mbox{
          \includegraphics[width=1. \columnwidth]{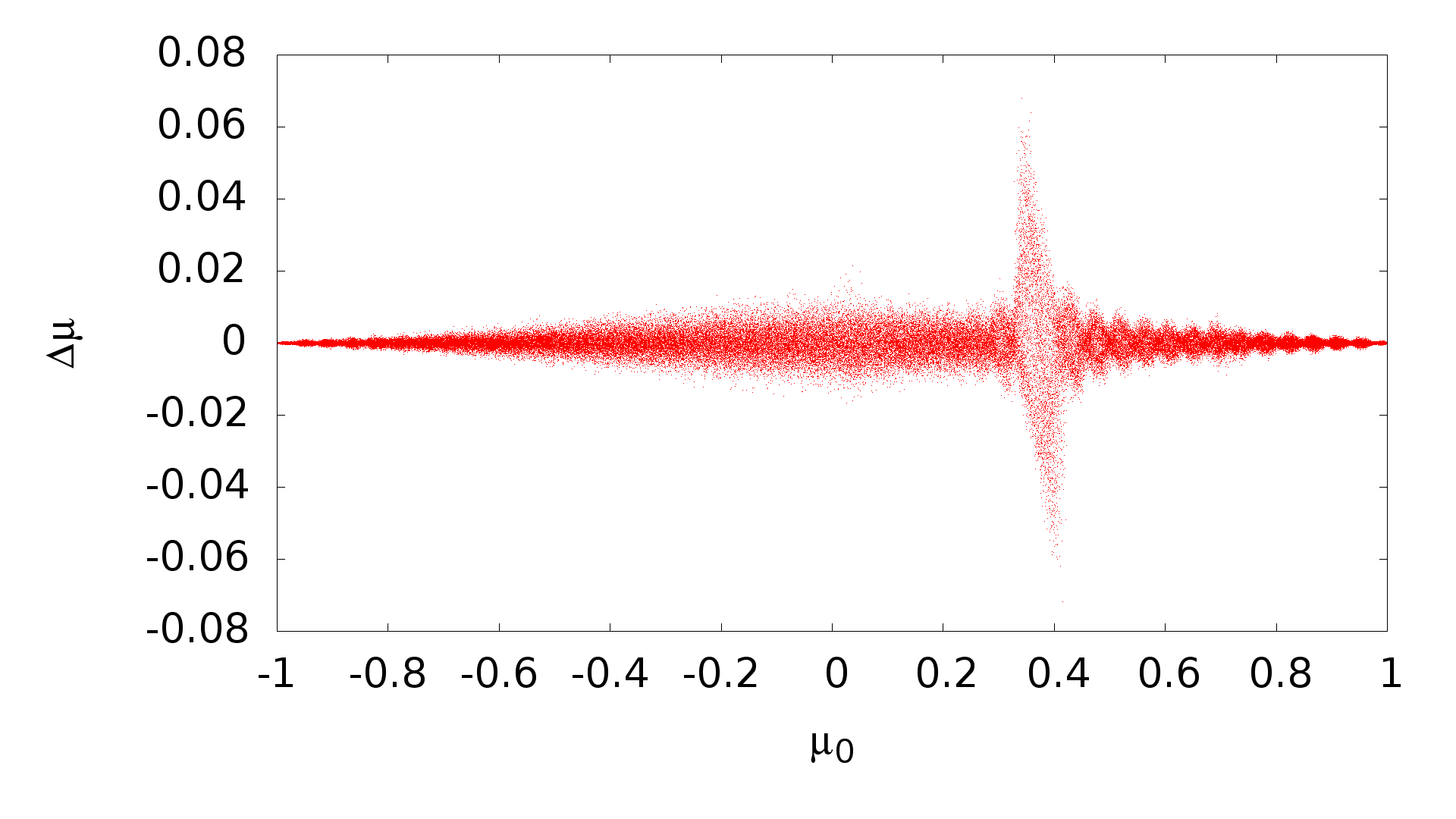}
         }
    \caption{Scatter plots for the high $B_0^h, \; 256^3$ gridsize, peak  $k'_\parallel = 2 \pi \cdot 24$, decay stage, $t=30$ gyration periods. The $n=1$ resonance is dominant, since the energy of the peaked wave mode remained at the parallel axis. Nevertheless, a small structure at $\mu_0=0.7$ indicates the $n=2$ resonance. The interactions with the left-handed circular polarised modes vanished after the driven stage and are not visible during the decay anymore.}
\label{fig:v35-small24-decay-deltamue-30gyr}
  \end{center}
\end{figure}

The last set of simulations used an increased magnetic background field $B_0$. Thus, the $\delta B / B_0$ ratio is decreased by an order of magnitude, which is
more consistent with the assumptions of the QLT. This claim is supported by Figs. \ref{fig:v35-small8-decay-deltamue-50gyr} and 
\ref{fig:v35-small24-decay-deltamue-30gyr} where each resonance has a reduced effective tilt because of the smaller scattering frequency and represents a dominant structure.
We note that the evolution time is roughly twice as long as within the weaker magnetic background field. 
A strong Cherenkov resonance is observed for the $k'_\parallel = 2 \pi \cdot 8$ peak. 
Also both $n=|1|$ resonances are visible. The small increase of the scattering rate at $0.53$ could be generated
by a resonant interaction with the higher harmonic at $k'_\parallel = 2 \pi \cdot 16$, which is also a dominant wave mode in this stage (see Fig. \ref{fig:v35-spherplots}, left-hand frame).

For the $k'_\parallel = 2 \pi \cdot 24$ peak, the fundamental resonance $n=1$ is strongest. A higher order of resonance is just slightly visible at $\mu_R=0.70$. An indication of the decay of the left-handed modes is the absence of resonant interactions $n<0$. These are visible during the driving of the peak, but vanish in the decay stage.

\subsection{Comparison between SQLT and particle simulations}\label{sec:comparisonsqlt}

\begin{figure}[ht]
  \begin{center}
    \mbox{
          \includegraphics[width=1. \columnwidth]{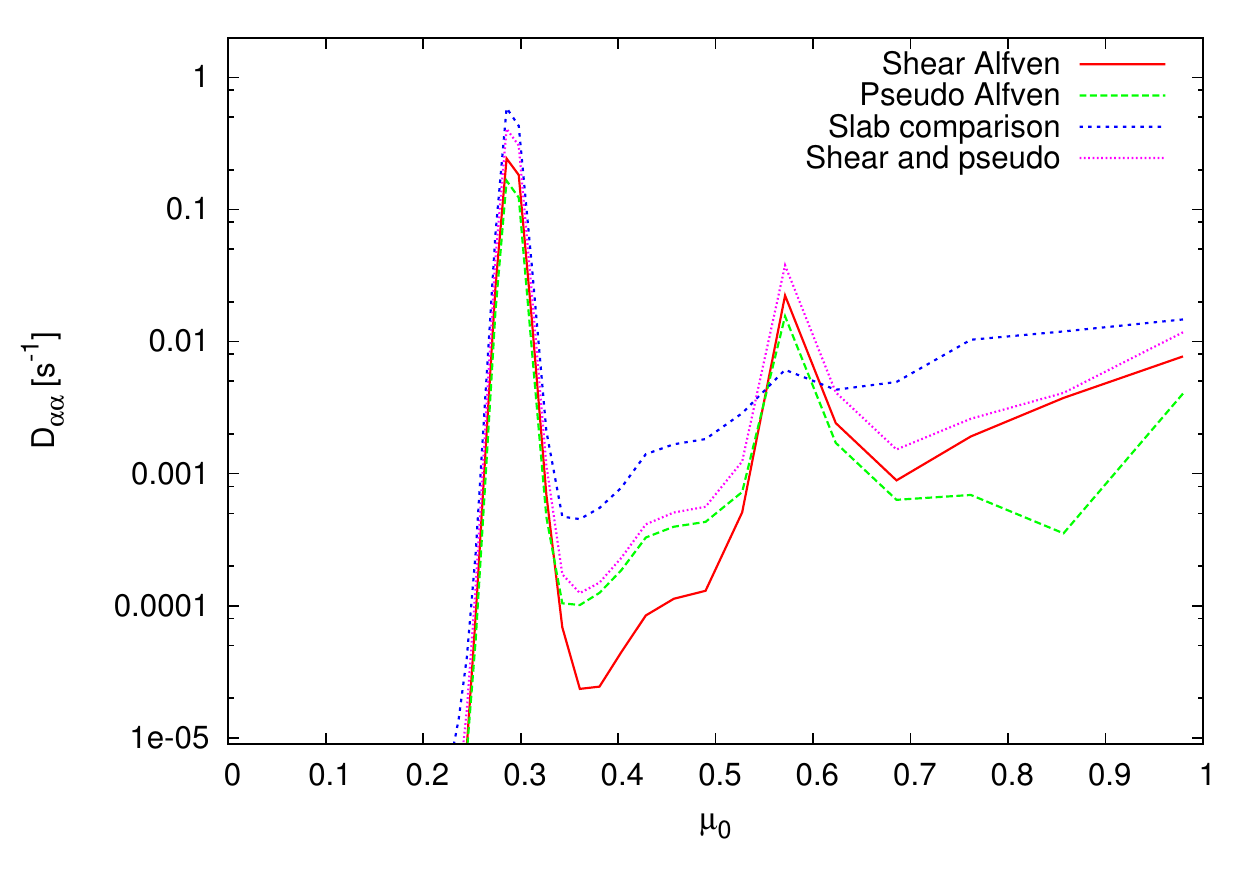}
         }
    \caption{Results for the magnetostatic SQLT approach for the $k'_\parallel = 2 \pi \cdot 24$ peak within the decay stage. The calculations were performed for each wave species separately. The slab approximation is comparable to the shear and pseudo Alfv\'en waves, but cannot reproduce the resonance at $\mu=0.58$.}
\label{fig:sqlt-v31-peak24-decay}
  \end{center}
\end{figure}

In the last section we present results of the SQLT approach and compare them to the particle simulations. For this purpose, the pitch-angle scattering
coefficient was calculated by using Eqs. (\ref{eq:dmm-sqltP}) and (\ref{eq:dmm-sqltA}) for each spectrum. 
In contrast to the particle simulations, the calculations via the SQLT approach produce a clear structure in the scattering coefficient, see Fig.  \ref{fig:sqlt-v31-peak24-decay}. 
That means that the resonances are not broadened by finite time and the effect of the tilt (see the toy model section) does not influence the values of $D_{\alpha \alpha}$.

As presented in Fig. \ref{fig:sqlt-v31-peak24-decay}, the magnetostatic calculations of the SQLT reproduce the resonant structures for $n=1,2$ very well.
However, the resonance gap for small pitch-angle cosine values causes a strong drop-off below $|\mu|\lesssim 0.25$. 
Furthermore, a slab approximation was made by the projection of the total wave energy to the parallel modes, i.e. each wave mode is assumed to be parallel.
The slab comparison gives a good approximation of the scattering coefficient, except for the resonant structures caused by the oblique waves ($|n|>1$). 
The scattering caused by the pseudo waves is comparable to the shear waves at the positions of the peaks. 
For values of $\mu = [0.35;0.55]$, the pseudo mode has higher scatter rates than the shear
mode, whose coefficient $D_{\alpha \alpha}$ is bigger elsewhere.
In the following, we restrict the presentation to the $D_{\alpha \alpha}$ sum of both wave modes, e.g. the fourth curve in Fig.\ref{fig:sqlt-v31-peak24-decay}.

\begin{figure}[ht]
  \begin{center}
    \mbox{
          \includegraphics[width=1. \columnwidth]{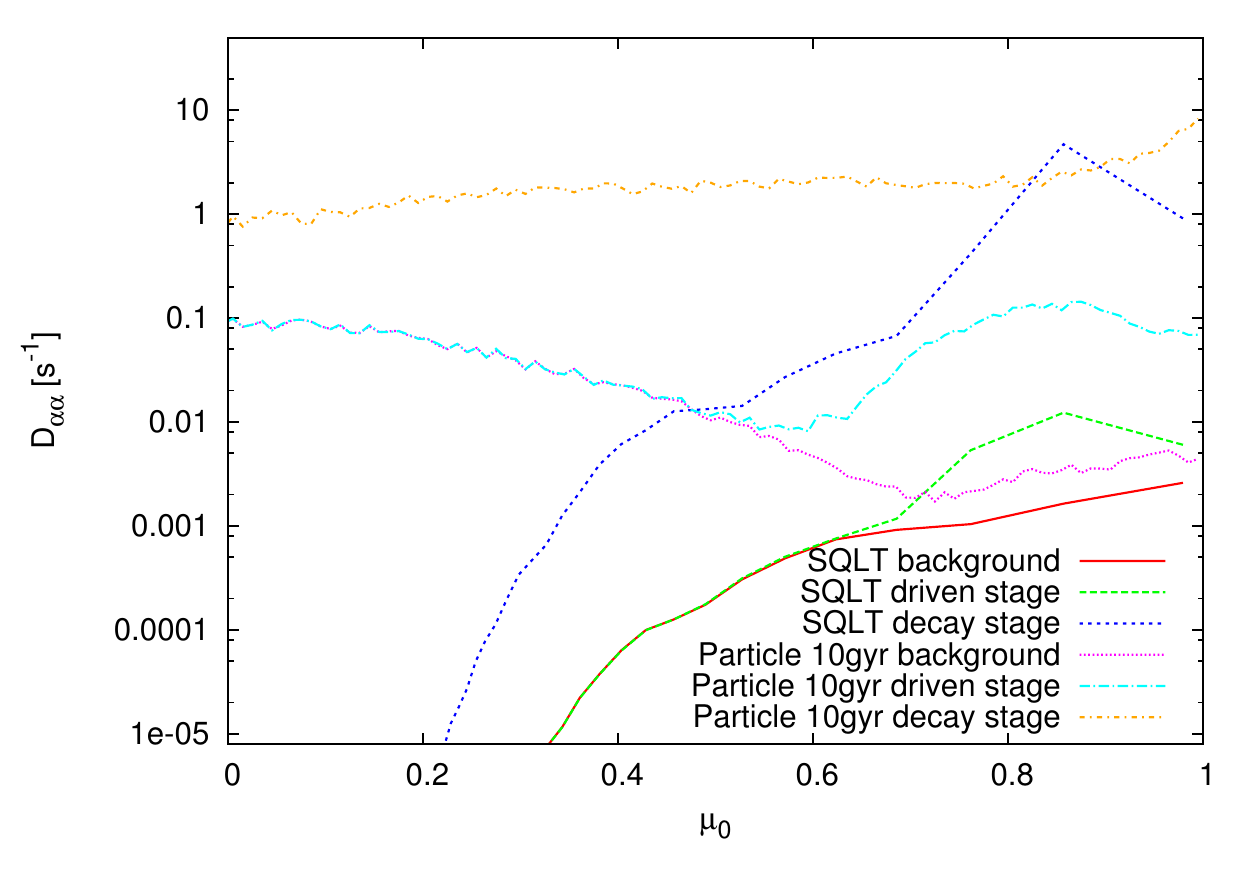}
         }
    \caption{Comparison between SQLT approach and particle simulation of the scattering coefficients $D_{\alpha \alpha}$ for the low $B_0^l, \; 256^3$ gridsize, peak $k'_\parallel = 2 \pi \cdot 8$. 
    For $\mu>0.6$, the SQLT coefficient is comparable to the particle simulations for both the background turbulence and the driven stage. 
    Due to the tilt (see Fig. \ref{fig:v31-small8-decay-deltamue}) the particles' $D_{\alpha \alpha}$ of the decay stage does not show a structure at the resonant $\mu_0$. Thus, the particle simulation curve does not match the SQLT model nicely.  }
\label{fig:v31-peak8-comparison-Daa}
  \end{center}
\end{figure}

As expected, the comparison between SQLT and particle simulation is best for the pure turbulent background case, as long as the Cherenkov resonance is small or
for higher values of $\mu$. Because of the resonance gap, it cannot be reproduced by the SQLT approach. 
However, the particle simulation presents a clear and broad interaction for $n=0$, as shown in Fig. \ref{fig:v31-peak8-comparison-Daa}.
Especially for the background and driven stage, the curves increase towards $\mu=0$, which is caused by the Cherenkov resonance. 
For a clearer interpretation of $D_{\alpha \alpha}$ it is again helpful to compare the $D_{\alpha \alpha}$ curve with the corresponding scatter plot.
According to the previous section, the $D_{\alpha \alpha}$ is not only increased at $\mu=0$ but also broadened by the tilt of the resonant peak. 
Nevertheless, the SQLT curve becomes comparable to the particle simulation for $\mu>0.6$. Because the scattering is concentrated at the purely parallel peak
during the driven stage, the scattering coefficient of the particle simulation shows the resonance very well and is also comparable to the SQLT calculation.
Even more complicated is the situation for the scattering at the peaked modes in the decay stage. The SQLT curve in Fig. \ref{fig:v31-peak8-comparison-Daa} shows the resonance at the predicted position $\mu_R=0.86$, whereas the $D_{\alpha \alpha}$ of the particle simulation seems to have no structure. 
This is again caused by the visible effect of the tilt at the resonances (see Fig. \ref{fig:v31-small8-decay-deltamue}).

\begin{figure}[ht]
  \begin{center}
    \mbox{
          \includegraphics[width=1. \columnwidth]{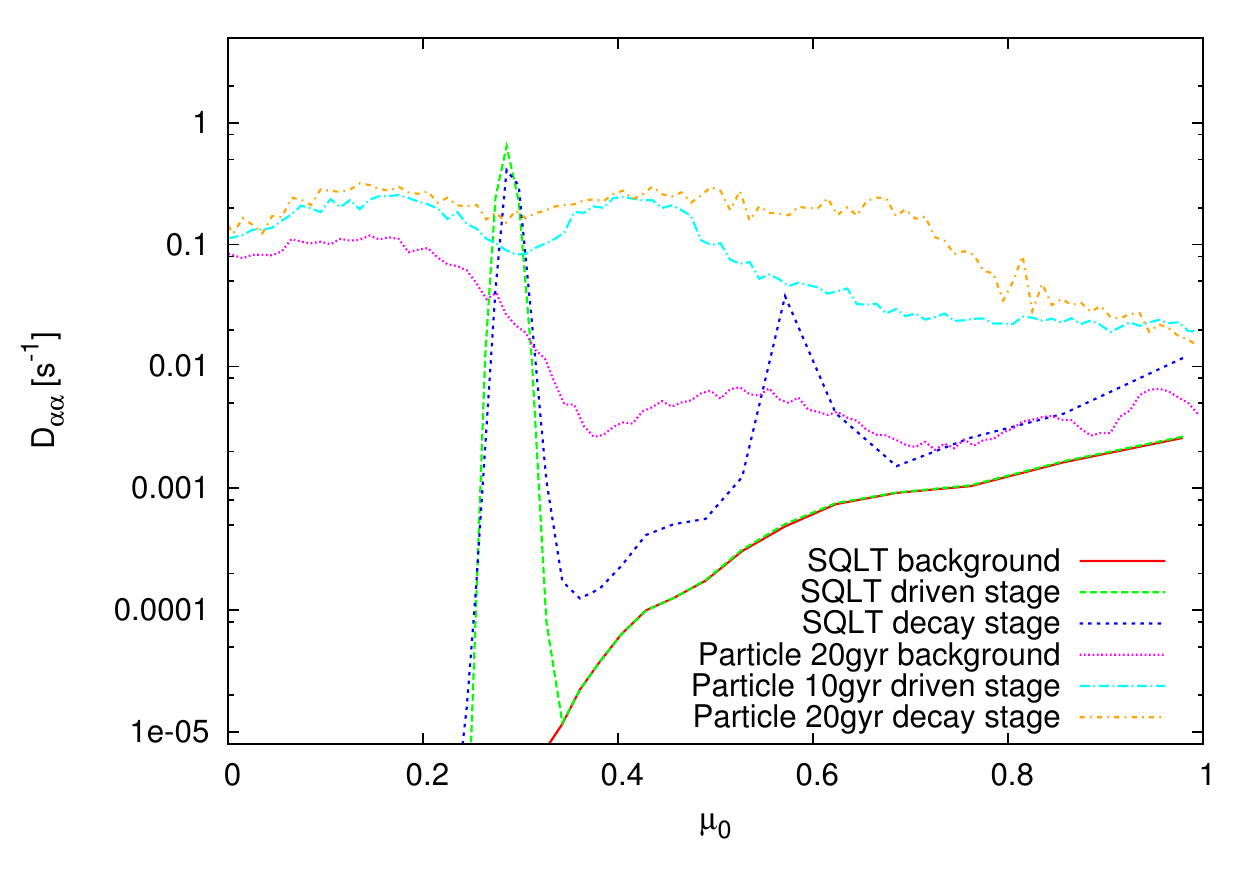}
         }
    \caption{Comparison between SQLT approach and particle simulation of the scattering coefficients $D_{\alpha \alpha}$ for the low $B_0^l, \; 256^3$ gridsize, peak position $k'_\parallel = 2 \pi \cdot 24$. The background turbulence is the same as in Fig. \ref{fig:v31-peak8-comparison-Daa}, but at $t=20$ gyr. The structure of the driven and decay stage of the particle $D_{\alpha \alpha}$ becomes clearer by comparison with Figs. \ref{fig:v31-small24-driven-deltamue} and \ref{fig:v31-small24-decay-deltamue}.}
\label{fig:v31-peak24-comparison-Daa}
  \end{center}
\end{figure}

Because the effective tilt within the $k'_\parallel = 2 \pi \cdot 24$ peak simulations is stronger and the resonance pattern for the decay stage is more
complex, the particle simulation results for $D_{\alpha \alpha}$ differ from the SQLT calculations. Nevertheless, the resonances in 
Fig. \ref{fig:v31-peak24-comparison-Daa} are located at the predicted positions.

\begin{figure}[ht]
  \begin{center}
    \mbox{
          \includegraphics[width=1. \columnwidth]{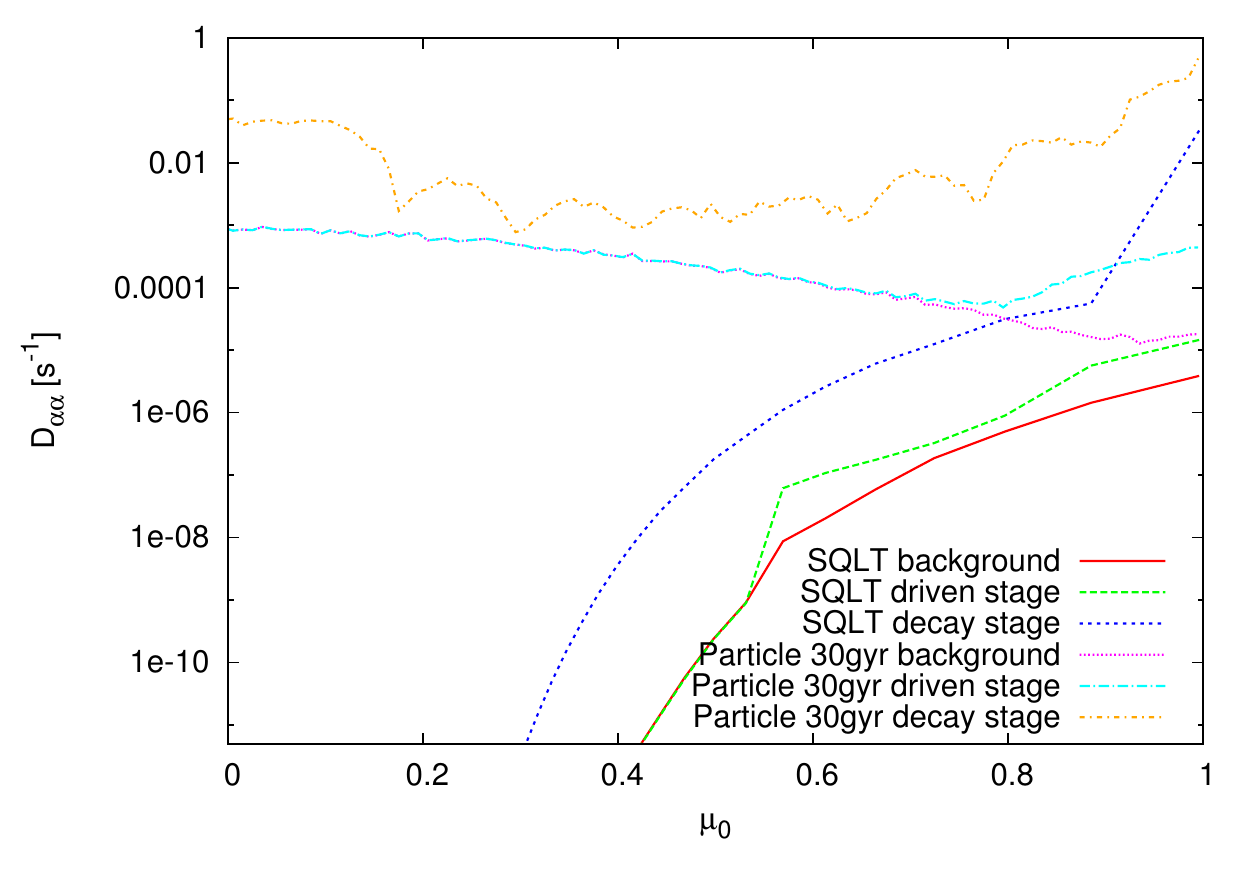}
         }
    \caption{Comparison between SQLT approach and particle simulation of the scattering coefficients $D_{\alpha \alpha}$ for the high $B_0^h, \; 256^3$ gridsize, peak position $k'_\parallel = 2 \pi \cdot 8$. The smaller $\delta B/B_0$ ratio fulfils the QLT approximation nicely. The particle $D_{\alpha \alpha}$ shows multiple maxima, where some of them are not necessarily resonances.}
\label{fig:v35-peak8-comparison-Daa}
  \end{center}
\end{figure}

\begin{figure}[ht]
  \begin{center}
    \mbox{
          \includegraphics[width=1. \columnwidth]{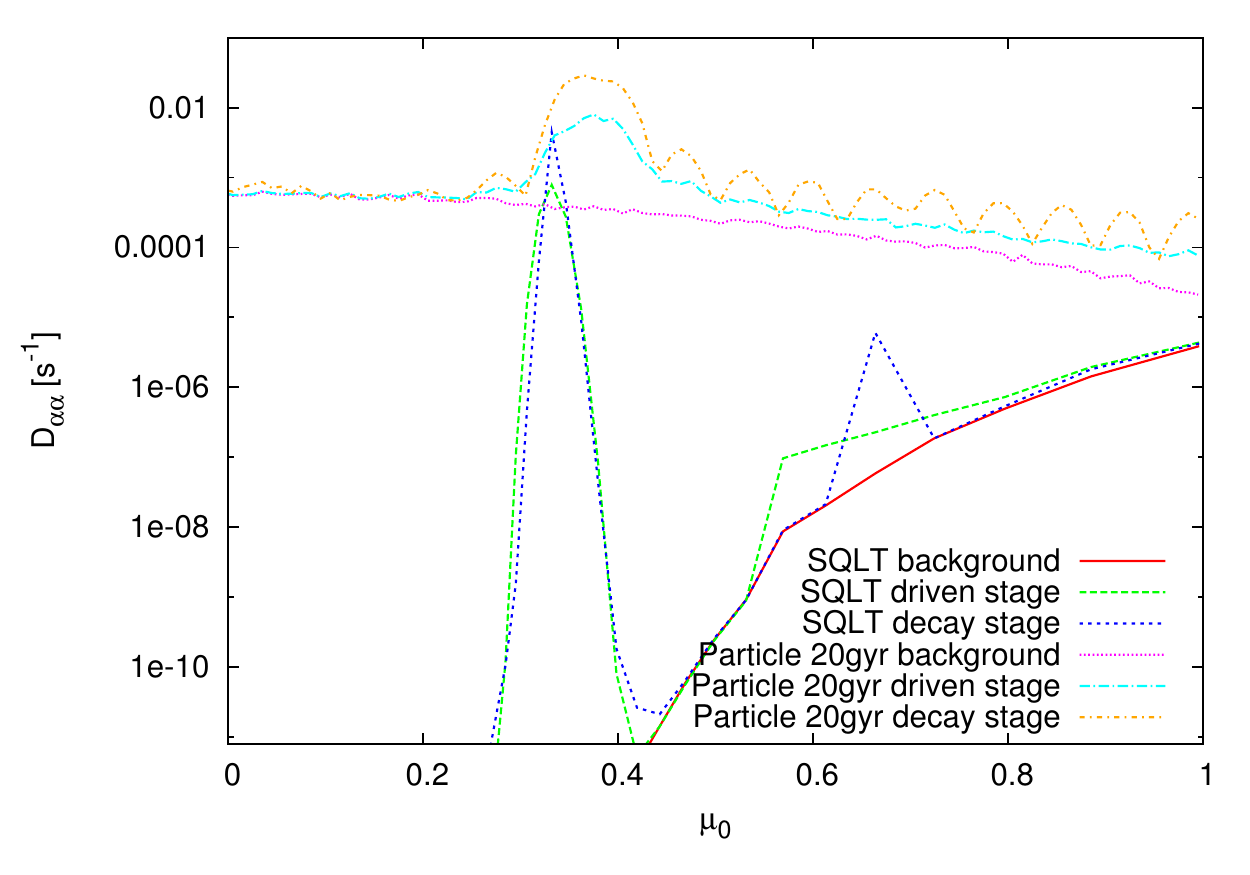}
         }
    \caption{Comparison between SQLT approach and particle simulation of the scattering coefficients $D_{\alpha \alpha}$ for the high $B_0^h, \; 256^3$ gridsize, peak $k'_\parallel = 2 \pi \cdot 24$. Both approaches show similar structure for the peaked modes. The shift of the resonance in SQLT can be explained by the magnetostatic assumption. The background turbulence does not match well because of the large resonance gap.}
\label{fig:v35-peak24-comparison-Daa}
  \end{center}
\end{figure}

The results of the simulation with a higher magnetic background field are presented in Figs. \ref{fig:v35-peak8-comparison-Daa} and  \ref{fig:v35-peak24-comparison-Daa}. 
The qualitative comparison to the SQLT is much better with the smaller $\delta B/B_0$ ratio. 
The particle scattering coefficients show clear resonant maxima at the predicted positions. A small shift of the SQLT resonances stems from the magnetostatic
assumption, which sets $\omega$ in Eq. (\ref{eq:mures}) to zero. This is valid for $\omega \ll \Omega$, which is not fulfilled for this simulation, 
since $\omega_8=177 \text{ s}^{-1}$, $\omega_{24}=532 \text{ s}^{-1}$ and $\Omega=4264\text{s}^{-1}$.  Thus the resonances are slightly shifted towards smaller
$\mu_R$, e.g. the $\mu_R=0.37$ for the $k'_\parallel = 2 \pi \cdot 24$ peak is $\mu_R=0.33$ in the magnetostatic limit. Another problem is the resonance gap,
which is significantly broader. Consequently, the SQLT curves cannot be compared quantitatively to the particle simulations.

\section{Conclusions}\label{sec:discussion}

To summarise, we firstly presented a simple wave--particle interaction model to achieve a fundamental understanding of the underlying processes and interpretation of numerical simulation results. 
The effects of tilted and shifted resonances were observed and further compared to QLT predictions.
Additionally, the difference between ballistic and resonant interactions could be shown in the context of time development.

Afterwards, we presented three different scenarios:
\begin{itemize}
 \item[1)] The first scenario used parameters to resemble conditions within a heliospheric plasma at three solar radii. 
 \item[2)] To study the influence of the resolution, the first simulation setup was used with a finer grid of $512^3$ cells (Appendix \ref{appendix:512runs}).
 \item[3)] The connection between QLT to a smaller $\delta B/B_0$ ratio was investigated by an approach using an artificially high $B_0$.
\end{itemize}
Each scenario was simulated with two different amplified wave modes with peak positions at $k'_\parallel = 2 \pi \cdot 8$ and 24.
Furthermore, two different evolution stages of the peak within the turbulence were investigated, which allows interactions with purely parallel peaked modes to be studied as well as oblique modes because of the peak development towards a perpendicular direction. 
All of the simulations were performed with $10^5$ protons. Their initial speed was chosen to be in resonance with the peaks. 
In addition, every turbulence spectrum was used for a magnetostatic QLT approach to calculate the pitch-angle scattering coefficient.
We presented resonance patterns with different methods: the scatter plot $\Delta \mu \, (\mu_0)$ and the scattering coefficient $D_{\alpha \alpha}\,(\mu_0)$ for both the numerical and semi-analytical quasilinear approach.

Our results show a good agreement between hybrid MHD particle simulations and QLT calculations for the scattering coefficient $D_{\alpha \alpha}$ for a turbulent broad--band spectrum. 
The comparison reveals that the SQLT calculations based on the MHD spectrum are missing the Cherenkov resonance ($n=0$), 
which is not covered by the QLT calculations used because quasilinear theory produces a singularity at $\mu=0$. 
Furthermore, the resonance gap at small values of $|\mu|$ is caused by the power spectrum, which does not have sufficient energy at high wave numbers within the dissipation range of the turbulence.
The direct particle simulations, on the other hand, show problems linked to broadened resonances: 
Because of the finite simulation time, the wave--particle resonance peaks are broadened, which in turn yields a
significant scattering background. 
With increasing simulation time, especially in simulations with higher $\delta B/B_0$, the resonances did not become narrow
but scattering would start to randomise.
Consequently, the assumption of $\delta$-shaped resonances does not apply to real particle scattering. 
Furthermore, the tilt in $\Delta \mu$, as shown in the scatter plots (e.g. Fig. \ref{fig:v31-small24-driven-deltamue}), causes spreading of the resonant peaks
over an interval of $|\mu|$ for the particle simulations.

Due to the tilt and the broadening of the resonances, the application of the scattering coefficient resulting from Eq. (\ref{eq:daa-coeff}) is questionable. 
The   $D_{\alpha \alpha}$ results are either difficult to interpret or without any reasonable structure. 
Consequently, the QLT approach does not compare well to the particle simulation results. 
We point out that in these cases a very nice tool in numerical simulations is the scatter plot. By evaluating the total change
of $\mu$ versus its initial state $\mu_0$, even small resonance patterns become visible. This can be used for correct interpretation of the scattering
coefficients.  Especially in cases with multiple overlapping resonances (e.g. Fig. \ref{fig:v31-small24-decay-deltamue}), the scatter plot yields important information. It should be noted, however, that the analysis of particle-scattering data requires more in-depth research.

A nice comparison between QLT and particle simulations is achieved with smaller $\delta B/B_0$ ratio of the turbulence and the peaked modes. 
This is not unexpected, because the assumption of unperturbed orbits is more reasonable. 
In this case the effective tilt in the scatter plots decreases significantly and $D_{\alpha \alpha}$ becomes more structured. Unfortunately, the resonance gap in the QLT approach expands in these scenarios, which leads to smaller absolute values of the scattering coefficient. 
This large gap stems from the used spectrum, where the magnetic background field was increased, leading to less resonant waves.

We conclude that for realistic particle transport, especially around $\mu \approx 0$, the QLT approach does not yield physical scattering coefficients. This is not unexpected, as it has been discussed in literature already \citepads{2004ApJ...616..617S}. 
We are not aware of any self-consistent theory describing the numerical particle results derived from our hybrid simulations. Further investigations are needed to disentangle the different transport processes involved in order to develop a new kind of transport theory.
We also note that even our numerical approach has limited validity for high $\delta B/B_0$ ratios. We will discuss more sophisticated numerical methods for transport analysis in a forthcoming paper.

\begin{acknowledgements}

We acknowledge support from Deutsche Forschungsgemeinschaft through grant SP 1124/3.\\
SL additionally acknowledges support from European Framework Program 7 Grant Agreement SEPServer - 262773 and the computing time at H\"ochstleistungs Rechenzentrum Stuttgart (HLRS).\\
TL acknowledges support from the UK Science and Technology Facilities Council (STFC) (grant ST/J001341/1) \\

\end{acknowledgements}

\appendix
\section{Time dependency of the pitch-angle cosine}
\label{appendix:deltamu}

The interaction of a charged particle with the fluctuation $\delta B$ of an Alfv\'en wave forces the particle to leave its gyro orbit. This procedure is referred to pitch-angle scattering. Consequently the change of parallel momentum is connected to this process via $p \totdiff{\mu}
{t} = \totdiff{p_\parallel}{t}$. Then the time derivative of $p_\parallel$ is given by the Lorentz force
\begin{align}
 \totdiff{p_\parallel}{t} = \frac{e c}{2 E_p} \sum_\pm(\mp i) p_{\perp} \delta B_{l,r},
\end{align}
where $E_p$ is the energy of the particle, which is assumed to be constant as the scattering is purely elastic. Since the parallel momentum remains constant during an unperturbed gyration, only the perpendicular momentum changes due to the circular movement $p_\perp = p_x \pm i p_y = p_{\perp 0} \exp[\mp i (\phi_0-\Omega t) ]$. This leads to a total change of $\mu$ for a particle with intial phase $\phi_0$ to the magnetic field with
\begin{align}
 \totdiff{\mu}{t} = \frac{e c}{2 p E_p} \sum_\pm(\mp i) p_{\perp 0} \exp[\mp i (\phi_0-\Omega t)] \delta B_{l,r}(\vec r), \label{eq:mudot}
\end{align}
\citepads{1974RvGSP..12..671L}. A similar approach is given by \citetads{schlickibook} using Eq. \ref{eq:sphaerischeVlasovgemittelt}. 
In this case the solution along the characteristics of the generalised force term $\bar g_\mu = \dot \mu$ is given by Eq. (12.2.4b) \citetads{schlickibook}. This result reduces to Eq. \ref{eq:mudot} by using the magnetostatic approximation, which assumes the electric field fluctuations to be negligible, $\delta E =0$. The coordinates used are still Eqs. \ref{eq:particlecoords}.
The fluctuation $\delta B$, i.e. of an Alfv\'en wave as used in the presented case, is given in Fourier space by
\begin{align}
  \delta B_{l,r}(\vec r)  = \frac{1}{8 \pi^3} \int \, \text{d}^3 \vec k \,\delta B_{l,r}(\vec k) \exp{(i \vec k \vec r)}.
\end{align}
The exponential function can be described by the generating Bessel functions $J_n$
\begin{align}
 \exp{(i \vec k \vec r)} = \sum_{n=-\infty}^\infty J_n(z)\exp{i(k_\parallel v_\parallel t + n (\Psi-\phi_0+\Omega t))},
 \end{align}
where the argument of $J_n$ is
\begin{align}
  z=\frac{k_\perp v \sqrt{1-\mu^2}}{\Omega}
\end{align}
and $\Psi=\cot^{-1}(k_x/k_y)$, $k_\perp=\sqrt{k_x^2+k_y^2}$.
The total time derivative of the pitch-angle cosine, separated in parallel and perpendicular interactions, reads then
\begin{align}
 \bar g_\mu = \totdiff{\mu}{t} =& \frac{e c}{2 p E_p} \sum_\pm(\mp i) p_{\perp 0} \exp[\mp i \phi_0] \times \nonumber \\
 &  \frac{1}{8 \pi^3} \int \, \text{d}^3 \vec k \; \delta B_{l,r}(\vec k) \sum_{n=-\infty}^\infty J_n\left(z\right) \times \nonumber \\
 &  \exp{[i n (\Psi - \phi_0) + i t (k_\parallel v_{\parallel} \pm \Omega - n \Omega)]},
\end{align}
For Alfv\'en waves $\delta B$ is aligned towards $\vec k \times \vec e_z B_0$ and consequently $\delta B_{l,r}(\vec k) = \delta B(\vec k) (\mp i)\exp(\pm i \Psi)$.
The time integration and the identity
\begin{align}
 \frac{2n}{z} J_n(z) = J_{n+1}(z) + J_{n-1}(z)
\end{align}
\citepads{1965hmfw.book.....A} then gives the time dependency of the pitch-angle cosine with
\begin{align}
 \Delta \mu (t) &= \mu(t) - \mu_0 = - \frac{e \, c \, p_{\perp 0}}{2 p E_p \; 8 \pi^3} \int \, \text{d}^3 \vec k \; \delta B(\vec k) \; \times \nonumber \\
 &\sum_{n=-\infty}^\infty  [J_{n+1}(z) + J_{n-1}(z)] \; \times  \nonumber \\
 & \exp{[i n (\Psi - \phi_0)]}
 \int_0^t \text{d} t' \exp{[i t' (k_\parallel v_{\parallel 0} - n \Omega)]}. \label{eq:mustreuung}
\end{align}
The application of a purely parallel propagating wave, as shown in our toy model, will then simplify this equation because \mbox{$z(k_\perp=0)=0$}. Thus, all Bessel functions vanish, except $J_0(0)=1$. This is the case for $n=\pm 1$. Furthermore, the $\vec k$ integration reduces by the assumption of a single wave $\vec k = \delta(k_\parallel-k_0)\delta(k_\perp) \vec {\hat k}$, which leads to the presented form in section \ref{sec:toymodel}.

\section{Derivation of the pitch-angle diffusion coefficients}
\label{appendix:derivdmm}
In the simulations, the turbulence consists of Alfv\'en and pseudo Alfv\'en
waves, thus the pitch-angle diffusion coefficient must be calculated
separately for the two modes. The two modes are decomposed using the
method presented \citetads{marongold}. For the pitch-angle diffusion
coefficient for Alfv\'en waves, \citetads{Schlickeiser2002} gives
\begin{align}
 D_{\mu\mu,A}=\frac{2\Omega^{2}(1-\mu^{2})}{B_{0}^{2}} & \sum_{n=-\infty}^{\infty}  \int\, d^{3}k\,\mathcal{R}\left(\mathbf{k},\omega\right)\left[1-\frac{\mu\omega}{k_{\parallel}v}\right]^{2} \times \nonumber \\
 &\left[\frac{n\, J_{n}\left(v_{\perp}k_{\perp}/\Omega\right)}{v_{\perp}k_{\perp}/\Omega}\right]^{2}P_{xx,A}(\mathbf{k}),
\end{align}
where $\mu$, $v$ and $\Omega$ again are the particle's pitch-angle cosine,
speed, and gyrofrequency, $v_{\perp}=v\,\sqrt{1-\mu^{2}}$, $P_{xx,A}$
the $xx$-component of the Alfv\'en mode turbulence power spectrum tensor,
and $B_{0}$ the background magnetic field. For the resonance function
$\mathcal{R}\left(\mathbf{k},\omega\right)$ we assume no damping,
which gives the delta function $\mathcal{R}\left(\mathbf{k},\omega\right)=\pi\,\delta\left(k_{\parallel}v_{\parallel}-\omega+n\Omega\right)$.
Thus, in the magnetostatic limit ($\omega=0$) we have
\begin{align}
 D_{\mu\mu,A}=\frac{2\Omega^{2}(1-\mu^{2})}{B_{0}^{2}} &\sum_{n=-\infty}^{\infty}\int\, d^{3}k\,\pi\delta\left(k_{\parallel}v_{\parallel}+n\Omega\right) \times \nonumber \\
 & \left[\frac{n\, J_{n}\left(v_{\perp}k_{\perp}/\Omega\right)}{v_{\perp}k_{\perp}/\Omega}\right]^{2}P_{xx,A}(\mathbf{k}).
\end{align}
 For the magnetosonic waves, \citetads{Schlickeiser2002} gives for the
fast mode on the cold plasma limit
\begin{align}
 D_{\mu\mu,M}\approx\frac{2\Omega^{2}(1-\mu^{2})}{B_{0}^{2}}& \sum_{n=-\infty}^{\infty} \int\, d^{3}k\,\mathcal{R}(\vec{k},\omega) \times \nonumber \\
 &\left[J_{n}'(v_{\perp}k_{\perp}/\Omega\right]^{2}P_{xx,M}(\mathbf{k})
\end{align}
for high particle velocities, $v_A\ll v$. As the polarisation of
the fast mode on the cold plasma limit is the same as for the pseudo Alfv\'en wave,
it is straightforward to show that the same form also holds for the pseudo Alfv\'en waves, with the
difference in the dispersion relation. However, by using the magnetostatic limit with $\omega = 0$ and again assuming no damping,
we get for the resonance function $\mathcal{R}(\vec{k},\omega)=\pi\delta\left(k_{\parallel}v_{\parallel}+\Omega\right),$
thus giving for the diffusion coefficient

\begin{align}
D_{\mu\mu,P}\approx\frac{2\Omega^{2}(1-\mu^{2})}{B_{0}^{2}} &\sum_{n=-\infty}^{\infty}\int\, d^{3}k\,\pi\delta\left(k_{\parallel}v_{\parallel}+n\Omega\right) \times \nonumber \\
&\left[J_{n}'(v_{\perp}k_{\perp}/\Omega)\right]^{2}P_{xx,P}(\vec{k}).
\end{align}

% Unlike for the Alfv\'en waves, the Cherenkov resonance, \mbox{$n=0$}, is nonzero
% for the Pseudo-Alfv\'en waves, and has to be considered separately.
% Unfortunately, the resonance function which is
% \begin{align}
%  \mathcal{R}_{n=0} = \frac{\pi\delta\left(v_{\parallel}-v_A\right)}{k_{\parallel}}
% \end{align}
% for this case, leads to a $D_{\mu\mu,P}$ Cherenkov term with a bare delta function.
% Since this is not physical, the SQLT model without any corrections does not support the Cherenkov resonance.

Unlike for the Alfv\'en waves, the Cherenkov resonance, \mbox{$n=0$}, is nonzero
for the pseudo Alfv\'en waves and has to be considered separately.
In this case, the resonance function is
\begin{align}
 \mathcal{R}_{n=0} = \frac{\pi\delta\left(v_{\parallel}\right)}{k_{\parallel}}.
\end{align}
This results in
\begin{align}
 D_{\mu\mu,P}^{n=0}=\frac{2\Omega^{2}(1-\mu^{2})}{B_{0}^{2}} & \delta\left(v_{\parallel}\right)  \times \nonumber \\
&\int\, d^{3}k\,\frac{\pi}{k_{\parallel}}\left[J_{0}'(v_{\perp}k_{\perp}/\Omega)\right]^{2}P_{xx,P}(\mathbf{k}),
\end{align}
which has a singularity at $v_{\parallel}=0$, and equals zero
elsewhere. Consequently, this term is not used by our model.

% The consequence of such a singularity is to remove the
% resonance gap between the pitch-angle hemispheres.

For other terms in the sum over $n$, we again use the magnetostatic
approximation $\omega=0$, thus giving the resonance condition
\begin{align}
\mathcal{R}(\vec{k},\omega)=\pi\delta\left(k_{\parallel}v_{\parallel}+n\Omega\right)
\end{align}
and the pitch-angle diffusion coefficient
\begin{align}
D_{\mu\mu,P} \approx  & \frac{2\Omega^{2}(1-\mu^{2})}{B_{0}^{2}} \sum_{n\ne0}\int\, d^{3}k\,\pi\delta\left(k_{\parallel}v_{\parallel}+n\Omega\right) \times \nonumber \\
&\left[J_{n}'(v_{\perp}k_{\perp}/\Omega)\right]^{2}P_{xx,P}(\mathbf{k}).
\end{align}

\section{Discretisation of the pitch-angle diffusion coefficients}
\label{appendix:discretdmm}
For the numerical calculation of the pitch-angle diffusion coefficient,
we consider the spectrum to be continuous in parallel direction, but
discrete in the perpendicular direction, i.e. $P_{xx}(k_{x,}k_{y},k_{z})=P_{xx}(h\Delta k,i\Delta k,k_{z})$,
with $h,\, i=-M,\ldots,M$. In this manner, the integrals over $k_{x}$~and~$k_{y}$
can be written as a sum, while the integral over~$k_{z}$ is evaluated
using the delta function. Then, the diffusion coefficient for the
Pseudo-Alfv\'en waves can be written as
\begin{align}
 D_{\mu\mu,P}^{n\neq0} &=\frac{2\pi\Omega^{2}(1-\mu^{2})}{B_{0}^{2}\left|v_{\parallel}\right|}  \times \nonumber \\
&\sum_{n=-\infty}^{\infty,n\neq0}\,\sum_{h,i=-M}^{M}\Delta k^{2}\left[J_{n}'(v_{\perp}k_{\perp}/\Omega)\right]^{2} P_{xx,P}(h\Delta k,i\Delta k,n\Omega/v_{\parallel}).
\end{align}
In this equation, $(n \Omega) /(\mu v)$ represents the parallel wavenumber at which $\delta(k_\parallel v_\parallel + n \Omega)$ is nonzero. Consequently, as turbulence data is available only at discrete wavenumbers, we define
$(n \Omega) /(\mu v) = l \Delta k$. Thus, with $l$ restricted to integer values, we find the values of $(v_\parallel = \mu v)$ at which the value of $D_{\mu \mu}$ can be solved.  
For a particle with $v=\Omega/(m\Delta k)$, with $m\le l\le M$, the pitch-angle is given as $\mu=m/l$.
Applying this discretisation, we have
\begin{align}
 &v_{\parallel}=v\,(m/l),  \nonumber \\
&v_{\perp}=v\,\sqrt{1-m^{2}/l^{2}}, \nonumber \\
\text{and}  \qquad &k_{\perp}=\sqrt{k_{x}^{2}+k_{y}^{2}}=\Delta k\,\sqrt{h^{2}+i^{2}}
\end{align}
and thus
\begin{align}
 D_{\mu\mu,P}^{n\neq0}\left(\frac{m}{l}\right) = & \; 2\pi\Omega\left(1-\frac{m^{2}}{l^{2}}\right)\, l  \; \times \nonumber \\
&\sum_{n=-M/l}^{M/l,\, n\neq0}\:\sum_{h,i=-M}^{M}\left[J_{n}'\left(\frac{1}{m}\sqrt{\left(1-\frac{m^{2}}{l^{2}}\right)\left(h^{2}+i^{2}\right)}\right)\right]^{2}  \times \nonumber \\
&\frac{\Delta k^{3}\: P_{xx,P}\left(h\Delta k,i\Delta k,n\, l\,\Delta k\right)}{B_{0}^{2}}.
\end{align}
For the Alfv\'en waves, using the same discretisation, we get
\begin{align}
 D_{\mu\mu,A}\left(\frac{m}{l}\right) = & \; 2\pi\Omega\left(1-\frac{m^{2}}{l^{2}}\right)\, l\; \times \nonumber \\
&\sum_{n=-M/l}^{M/l}\:\sum_{h,i=-M}^{M}\left[\frac{n\, J_{n}\left(\frac{1}{m}\sqrt{\left(1-\frac{m^{2}}{l^{2}}\right)\left(h^{2}+i^{2}\right)}\right)}{\frac{1}{m}\sqrt{\left(1-\frac{m^{2}}{l^{2}}\right)\left(h^{2}+i^{2}\right)}}\right]^{2}  \times \nonumber \\
&\frac{\Delta k^{3}\: P_{xx,A}\left(h\Delta k,i\Delta k,n\, l\,\Delta k\right)}{B_{0}^{2}}.
\end{align}

\section{$512^3$ Results}
\label{appendix:512runs}

In this section we present the results of the investigation of the resolution by using a spatial grid of $512^3$ cells. Again, the MHD simulations were performed, first for the background turbulence, afterward with the peaked modes at $k'_\parallel = 2 \pi \cdot 8$ and 24. The results are in conformance with the other setups. In particular, we observed the generation of higher harmonic wave modes again.

\begin{figure}[ht]
  \begin{center}
    \mbox{
          \includegraphics[width=1. \columnwidth]{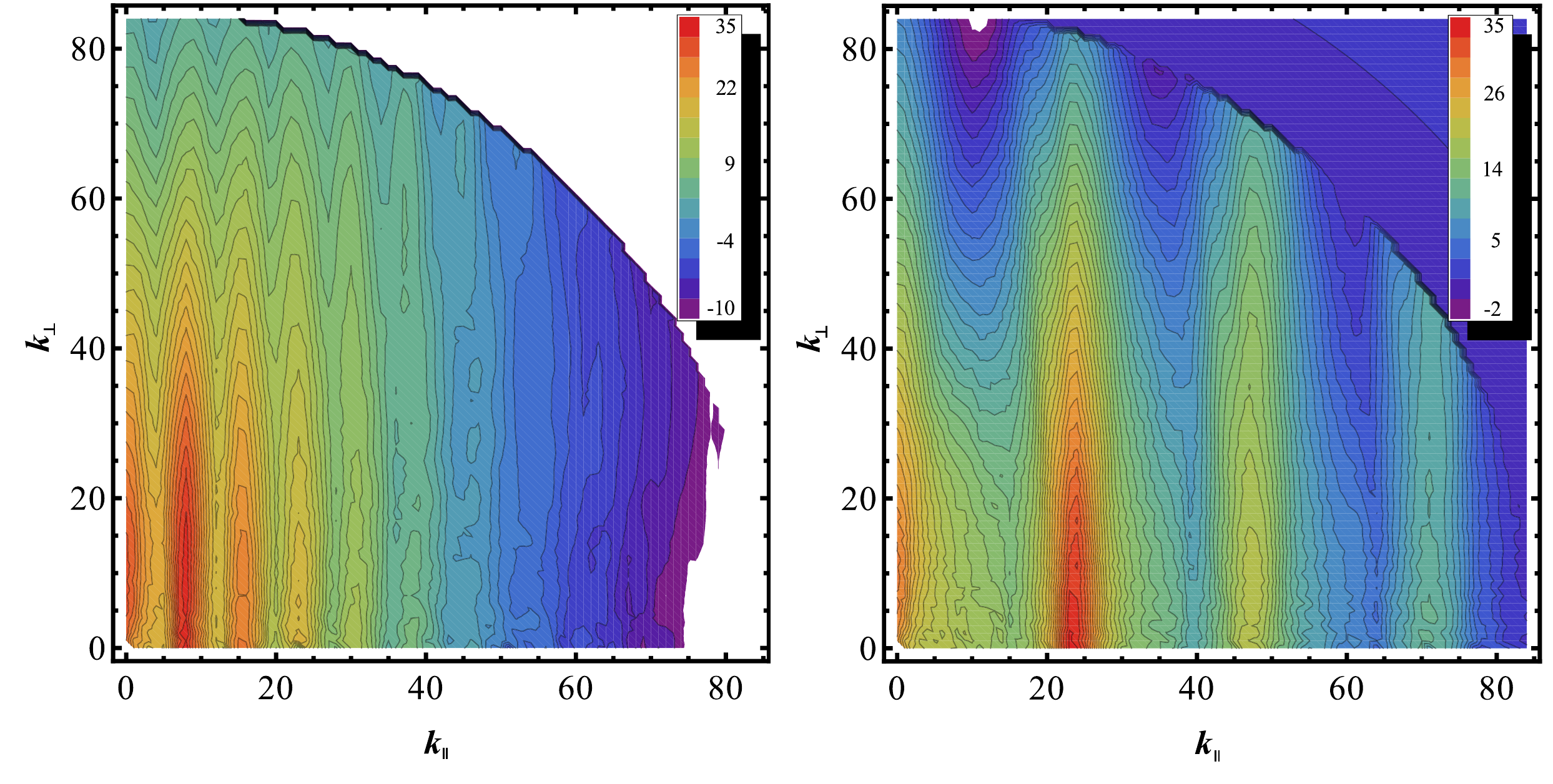}
         }
    \caption{Two-dimensional magnetic energy spectra of the decay stage for both peaks in the simulation with $B_0^l=0.174$ G and higher resolution with a grid of $512^3$ cells. The left figure shows the state for the peak at $k'_\parallel = 2 \pi \cdot 8$ at $t=14.45$ s. The right figure shows the $k'_\parallel = 2 \pi \cdot 24$ peak at $t=3.4$ s. The larger Fourier space grid reveals the higher harmonics of the $k'_\parallel = 2 \pi \cdot 24$ peak.}
\label{fig:512_8u24_decay-spherplots}
  \end{center}
\end{figure}

These harmonics are also visible for the peak at $k'_\parallel = 2 \pi \cdot 24$ within the higher resolved simulation with a $512^3$ grid. In this simulation setup the number of active modes is eight times greater, which means the antialiasing edge is shifted by a factor two to $k' = 2 \pi \cdot 86$. Consequently, the higher harmonics at $k'_\parallel = 2 \pi \cdot 48$ and 72 are visible. The evolution of the peaks in both simulations is comparable to the $256^3$ grid simulations. A dominant energy transport towards high perpendicular wavenumbers is observed.

\begin{figure}[ht]
  \begin{center}
    \mbox{
          \includegraphics[width=1. \columnwidth]{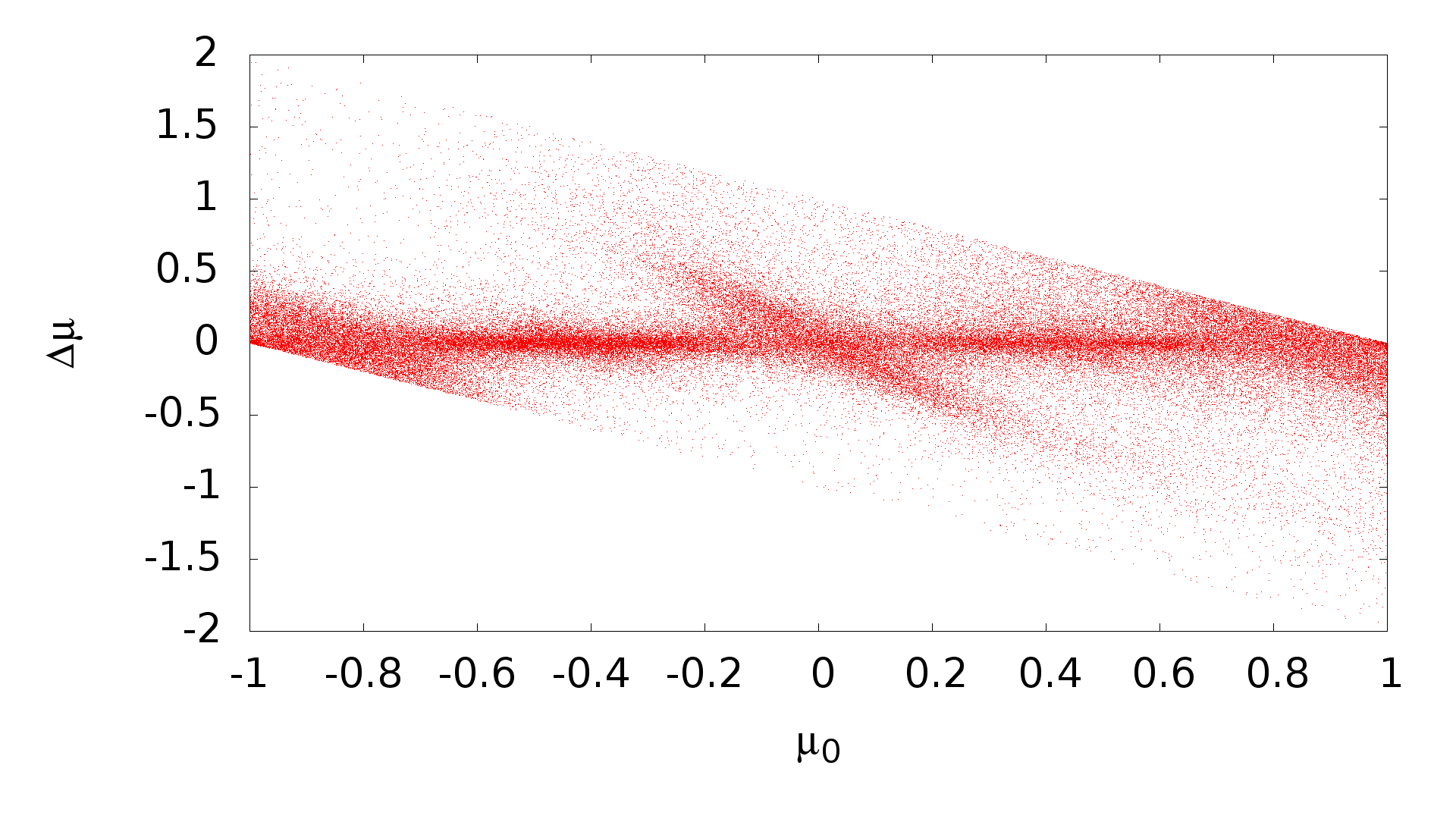}
         }
    \caption{Scatter plots for the low $B_0^l, \; 512^3$ gridsize, peak position $k'_\parallel = 2 \pi \cdot 8$, decay stage, $t=25$ gyration periods. The higher resolution of the grid causes slightly stronger scattering, due to the higher amount of active modes. The resonance patterns are comparable to the $256^3$ grid.}
\label{fig:v31-512-small8-decay-deltamue-25gyr}
  \end{center}
\end{figure}

The test particle simulations led to comparable resonance patterns. As presented in Fig. \ref{fig:v31-512-small8-decay-deltamue-25gyr}
the resonant interactions at $\mu_R = 0, 1$ and $-0.95$ are strongly tilted and a significant amount of particles reach the maximum $\Delta \mu$, as indicated by the sharp thresholds. 
The scattering coefficient is in this case again without any structure and hence not shown here. The stronger scattering is primarily
caused by the higher wave modes, which are not truncated by the antialiasing anymore and hence contribute to the wave--particle interactions. The test particle simulations were performed for the decay stage of the peaks only because of their huge computational effort.

\begin{figure}[ht]
  \begin{center}
    \mbox{
          \includegraphics[width=1. \columnwidth]{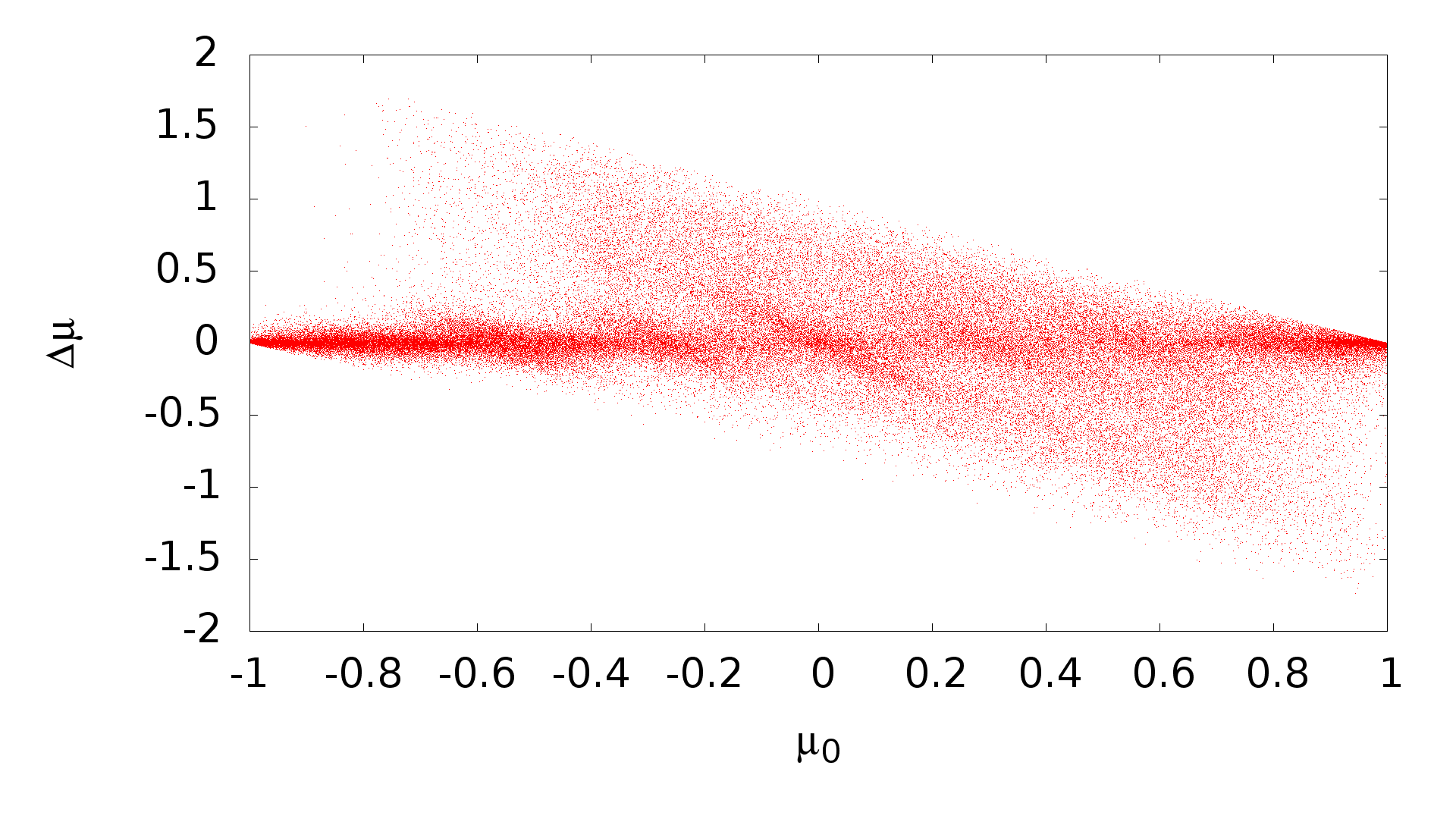}
         }
    \caption{Scatter plots for the low $B_0^l, \; 512^3$ gridsize, peak position $k'_\parallel = 2 \pi \cdot 24$, decay stage, $t=25$ gyration periods. Because of the higher resolution, higher harmonics of the $k'_\parallel = 2 \pi \cdot 24$ peak have also developed and interact with the particle. This leads to more diffuse scattering. The resonances are barely visible.}
\label{fig:v31-512-small24-decay-deltamue-25gyr}
  \end{center}
\end{figure}

Additionally, an increase of the scattering rate at the $k'_\parallel = 2 \pi \cdot 24$ peak within the $512^3$ grid was observed. Consequently, the resonances are not as significant as in the smaller grid. When comparing Fig. \ref{fig:v31-small24-decay-deltamue} and \ref{fig:v31-512-small24-decay-deltamue-25gyr} it is harder to recognize the resonant structures.

\begin{figure}[ht]
  \begin{center}
    \mbox{
          \includegraphics[width=1. \columnwidth]{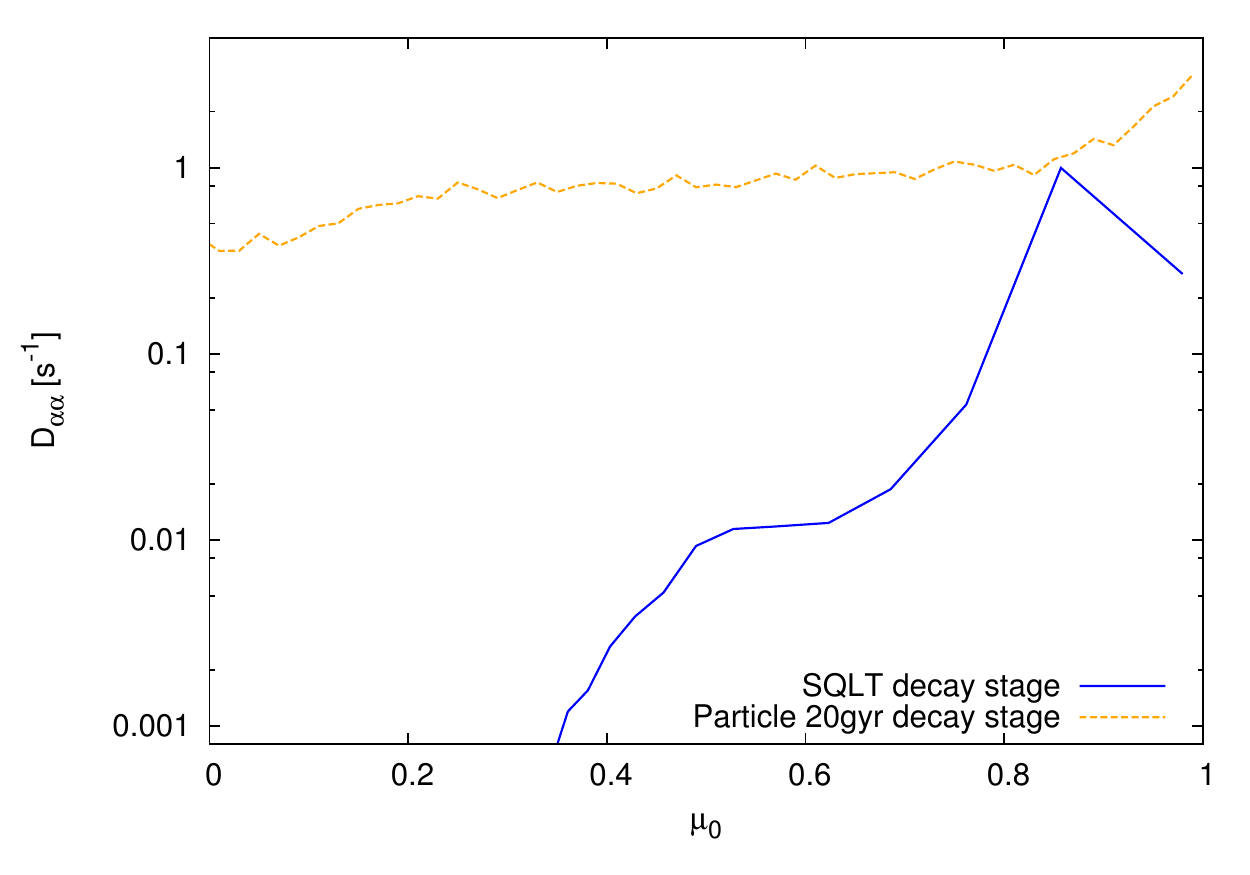}
         }
    \caption{Comparison between SQLT approach and particle simulation of the scattering coefficients $D_{\alpha \alpha}$ for the low $B_0^l, \; 512^3$ gridsize, peak position $k'_\parallel = 2 \pi \cdot 8$. The results are similar to Fig. \ref{fig:v31-peak8-comparison-Daa}.}
\label{fig:v31-512-peak8-comparison-Daa}
  \end{center}
\end{figure}

\begin{figure}[ht]
  \begin{center}
    \mbox{
          \includegraphics[width=1. \columnwidth]{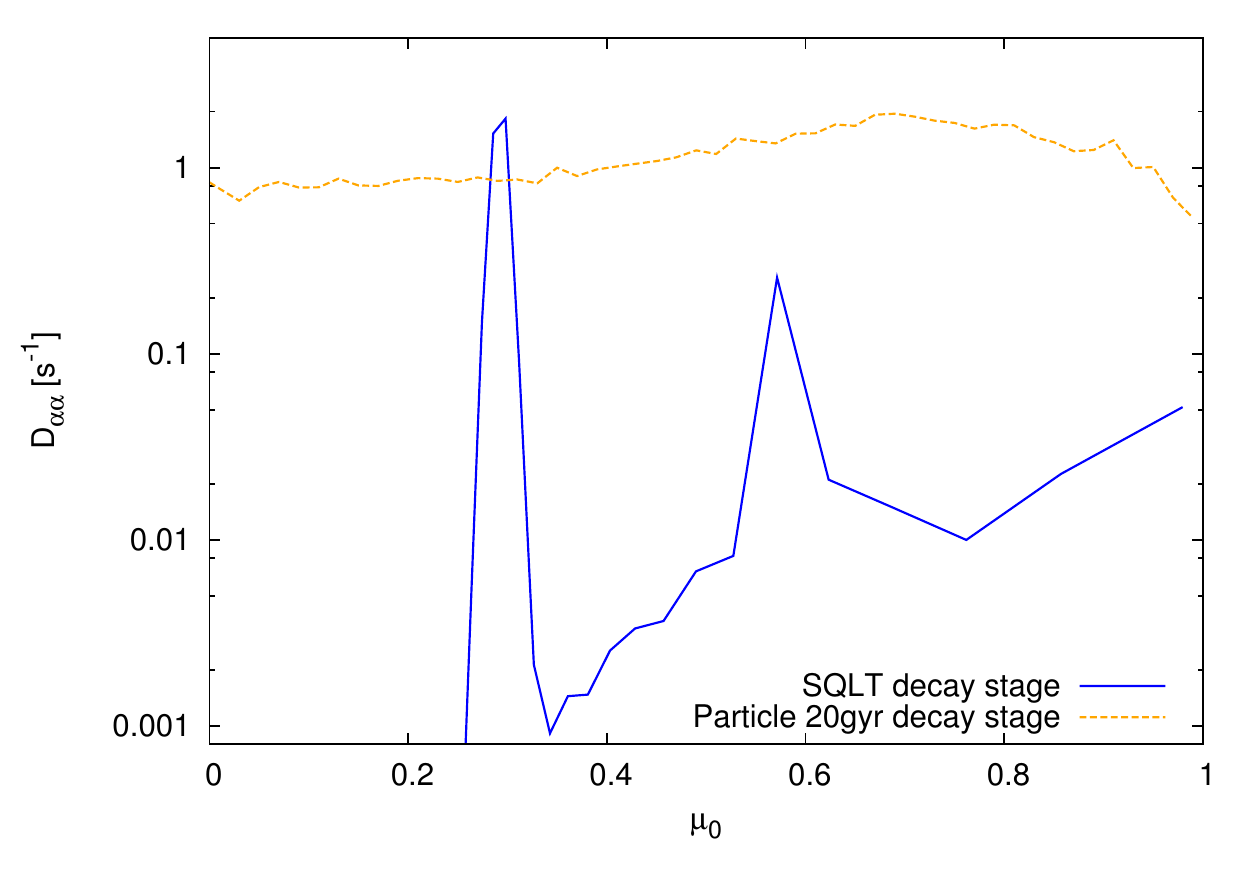}
         }
    \caption{Comparison between SQLT approach and particle simulation of the scattering coefficients $D_{\alpha \alpha}$ for the low $B_0^l, \; 512^3$ gridsize, peak position $k'_\parallel = 2 \pi \cdot 24$. Because of the increased active modes the scattering became stronger and less structured (see Fig. \ref{fig:v31-512-small24-decay-deltamue-25gyr}). Consequently, the $D_{\alpha \alpha}$ of the simulation cannot be compared well to the QLT approach.}
\label{fig:v31-512-peak24-comparison-Daa}
  \end{center}
\end{figure}

As discussed previously, the $512^3$ gridsize particle simulation within the decay stage led to strong effectively tilted resonances and consequently a very unstructured $D_{\alpha \alpha}$ curve. This can be observed in both Figs. \ref{fig:v31-512-peak8-comparison-Daa} and \ref{fig:v31-512-peak24-comparison-Daa}. Thus, the results differ greatly from those given by SQLT.

\bibliographystyle{aa}
\bibliography{ref}

\end{document}